\documentclass[12pt]{iopart}
\usepackage{iopams}  
\usepackage{bm}
\usepackage{mathrsfs}
\usepackage[utf8]{inputenc} 
\usepackage{enumerate}
\usepackage{appendix}
\usepackage{dcolumn}
\usepackage{multirow}
\usepackage{float}
\usepackage{color}
\usepackage{cancel}
\usepackage[utf8]{inputenc}
\usepackage{hyperref}

\makeatletter
\long\def\@makefntext#1{\parindent 1em\noindent 
 \makebox[1em][l]{\footnotesize\rm$\m@th{\arabic{footnote}}$}%
 \footnotesize\rm #1}
\def\@makefnmark{\hbox{${\arabic{footnote}}\m@th$}}
\def\@thefnmark{\arabic{footnote}}
\makeatother

\begin{document}
\title[The information loss problem: an analogue gravity perspective]
{The information loss problem: an analogue gravity perspective}

\author{Stefano Liberati,$^{1,2,3}$ Giovanni Tricella,$^{1,2,3}$ and Andrea Trombettoni$^{1,2}$}

\address{$^1$ SISSA - International School for Advanced Studies, via Bonomea 265, \\
\qquad 34136 Trieste, Italy.}
\address{$^2$ INFN Sezione di Trieste, via Valerio 2, Trieste, Italy.}
\address{$^3$ IFPU - Institute for Fundamental Physics of the Universe, Via Beirut 2, \\
\qquad 34014 Trieste, Italy.}
\ead{liberati@sissa.it, gtricell@sissa.it, andreatr@sissa.it}
\vspace{10pt}

%
%

\definecolor{purple}{rgb}{1,0,1}
\newcommand{\red}[1]{{\slshape\color{red} #1}}
\newcommand{\blue}[1]{{\slshape\color{blue} #1}}
\newcommand{\purple}[1]{{\slshape\color{purple} #1}}

\begin{abstract}
Analogue gravity can be used to reproduce the phenomenology of quantum
field theory in curved spacetime and in particular phenomena such as
cosmological particle creation and Hawking radiation.
In black hole physics, taking into account the backreaction of such
effects on the metric requires an extension to semiclassical gravity
and leads to an apparent inconsistency in the theory: the black hole
evaporation induces a breakdown of the unitary quantum evolution
leading to the so called information loss problem. Here we show that 
analogue gravity can provide an interesting perspective on the resolution
of this problem, albeit the backreaction in analogue systems is not described 
by semiclassical Einstein equations.  In
particular, by looking at the simpler problem of cosmological particle
creation, we show, in the context of Bose--Einstein condensates analogue gravity, that the
emerging analogue geometry and quasi-particles have correlations due
to the quantum nature of the atomic degrees of freedom underlying the
emergent spacetime. The quantum evolution is, of course, always
unitary, but on the whole Hilbert space, which cannot be exactly factorised
\emph{a posteriori} in geometry and quasi-particle components.
In analogy, in a black hole evaporation one should expect a continuous
process creating correlations between the Hawking quanta and the
microscopic quantum degrees of freedom of spacetime, implying so that
only a full quantum gravity treatment would be able to resolve the
information loss problem by proving the unitary evolution on the full
Hilbert space.

\bigskip


\end{abstract}

\vspace{2pc}

\maketitle

\def\tr{{\mathrm{tr}}}
\def\cof{{\mathrm{cof}}}
\def\pdet{{\mathrm{pdet}}}
\def\d{{\mathrm{d}}}

%

\hrule
\tableofcontents
\markboth{The information loss problem: an analogue gravity perspective}{The information loss problem: an analogue gravity perspective}
\bigskip
\hrule
\bigskip

\section{Introduction}
Albeit being discovered more than 40 years ago, 
Hawking radiation is still at the center of much work in theoretical physics 
due to its puzzling features and its prominent role in connecting general 
relativity, quantum field theory and thermodynamics. Among the new 
themes stimulated by Hawking's discovery, two have emerged as 
most pressing: the so called transplanckian problem and the 
information loss problem.

The transplanckian problem stems from the fact that infrared 
Hawking quanta observed at late times at infinity seems 
to require the extension of relativistic quantum field theories 
in curved spacetime well within the UV completion of the theory, \emph{i.e.} 
the Hawking calculation seems to require a strong assumption about 
the structure of the theory at the Planck scale and beyond. 

It was mainly with this open issue in mind that in 1981 
Unruh introduced the idea to simulate in condensed matter systems 
black holes spacetime and the dynamics of fields above 
them~\cite{Unruh:1980cg}. Such analogue models of gravity 
are provided by several condensed-matter/optical systems 
in which the excitations propagate in an effectively relativistic fashion 
on an emergent pseudo-Riemannian geometry induced by the medium.
Indeed, analogue gravity have played a pivotal role in the past years by 
providing a test bench for many open issues in quantum field theory 
in curved spacetime and in demonstrating the robustness of Hawking radiation 
and cosmological primordial spectrum of perturbations stemming from 
inflation against possible UV completions of the theory 
(see \emph{e.g.}~\cite{Barcelo:2005fc} for a comprehensive review). 
More recently, the same models have offered a valuable framework within 
which current ideas about the emergence of spacetime and its dynamics could 
be discussed via convenient toy models~\cite{Visser:2001fe,Girelli:2008gc,Finazzi:2011zw,Belenchia:2014hga}.

Among the various analogue systems, a preeminent role has been 
played by Bose--Einstein condensates (BEC) because these are macroscopic 
quantum systems whose phonons/quasi-particle excitations can be 
meaningfully treated quantum mechanically. Hence they can be used to fully 
simulate the above mentioned quantum 
phenomena~\cite{Garay:1999sk, Garay:2000jj, Barcelo:2003wu} and also 
as an experimental test bench of these ideas~\cite{Lahav:2009wx,Steinhauer:2015ava,Steinhauer:2015saa,deNova:2018rld}.

In what follows, 
we shall argue that these systems can not only reproduce Hawking 
radiation and address the transplanckian problem, but they can also 
provide a precious insight into the information loss problem.
For gravitational black holes, the latter seems to be a direct consequence of the backreaction
of Hawking radiation, leading
to the decrease of the black hole mass and of the region enclosed by the horizon. 
The natural endpoint of such process into a complete evaporation of the object leads to
a thermal bath over a flat spacetime which appears to be incompatible
with a unitary evolution of the quantum fields from the initial state to the final one.~\footnote{We are not considering here alternative solutions
  such as long-living remnants, as these are as well problematic in other
  ways~\cite{Giddings1993a, Giddings1993b, Susskind1995, Chen2014, Hossenfelder2009}, or they imply deviations
  from the black hole structure at macroscopic scale, see \emph{e.g.}.~\cite{Almheiri2013}.}

Of course, the BEC system at the fundamental level cannot violate unitary evolution. 
However, it is easy to see that one can conceive analogue black holes provided with singular regions for the emergent spacetime where the description of quasi-particles propagating on an analogue geometry fails.~\footnote{For example, one can describe flows characterized by regions where the  hydrodynamical approximation fails even without necessarily having loss of atoms from the systems.}   In such cases, despite the full dynamics being unitary, it seems that a trace over the quasi-particle falling in these ``analogue singularities" would be necessary, so leading to an apparent loss of unitarity from the analogue system point of view. 
The scope of the present investigation is to see how such unitarity evolution is preserved on the full Hilbert space.

However, in addressing the information loss problem in gravity, 
the spacetime geometry and the quantum fields are implicitly assumed
to be separated sectors of the Hilbert space.  
In the BEC analogue this assumption is 
reflected in the approximation that the quantum nature of the operator 
$\hat{a}_{k=0}$, 
creating particles in the background condensate, 
can be neglected. 
Therefore, in standard analogue gravity a description of the quantum evolution on the full Hilbert space seems precluded.

Nonetheless, it is possible to retain the quantum nature of the condensate operator as well as to describe their possible correlations with quasi-particles within an improved Bogoliubov description, namely the number-conserving approach~\cite{Pethick20081}. Remarkably, we shall show that the analogue gravity framework can be extended also in this context. Using the simpler setting of a cosmological particle creation we shall describe how this entails the continuous generation of correlations between the condensate atoms and the quasi-particles. Such correlations are responsible for (and in turn consequence of) the non factorisability of the Hilbert space and are assuring in any circumstances the unitary evolution of the full system. The lesson to be drawn is that in gravitational systems only a full quantum gravity description could account for the mixing between gravitational and matter quantum degrees of freedom and resolve in this way the apparent paradoxes posed by black hole evaporation in quantum field theory on curved spacetime.

The paper is organised as follows. In Section~\ref{sec:AG-BEC} 
we briefly recall the analogue gravity model for
a non-relativistic BEC in mean-field approximation.
In Section~\ref{sec:orbitas} we review the time-dependent orbitals formalism
and how it is employed in the general characterisation of condensates and
in the description of the dynamics. In Section~\ref{QP-BEC} we introduce the
number-conserving formalism. In Section~\ref{sec:AG-Num} we discuss the 
conditions under which we can obtain an analogue gravity model in this
general case. Finally, we analyse in Section~\ref{sec:SC} how in analogue gravity
the quasi-particle dynamics affects the condensate and in Section~\ref{sec:SQSS} 
how the condensate and the excited part of the full state are entangled, 
showing that the unitarity of the evolution is a feature of the system considered in its entirety.
Section~\ref{sec:Conc} is devoted to a discussion of the obtained
results and of the perspectives opened by our findings.

\section{Analogue gravity}\label{sec:AG-BEC}
In this Section we briefly review how to realise 
a setup for analogue gravity~\cite{Barcelo:2005fc} with BEC 
in the Bogoliubov approximation, with a bosonic low-energy (non-relativistic) 
atomic system. In particular, we consider the simplest case, 
where the interaction potential is given by a local $2$-body interaction. 
The Hamiltonian operator, the equation of motion and commutation relations 
in second quantization formalism are as usual
\begin{eqnarray}
\fl
\qquad \qquad \quad H&=&\int\mathrm{d}x\left[\phi^{\dagger}\left(x\right)\left(-\frac{\nabla^{2}}{2m}\phi\left(x\right)\right)+\frac{\lambda}{2}\phi^{\dagger}\left(x\right)\phi^{\dagger}\left(x\right)\phi\left(x\right)\phi\left(x\right)\right]\, ,\label{MB Hamiltonian}\\
\fl
\qquad i\partial_{t}\phi\left(x\right)&=&\left[\phi\left(x\right),H\right]=-\frac{\nabla^{2}}{2m}\phi\left(x\right)+\lambda\phi^{\dagger}\left(x\right)\phi\left(x\right)\phi\left(x\right)\, , \label{Full dynamics}\\
\fl
\left[\phi\left(x\right),\phi^{\dagger}\left(y\right)\right]&=&\delta\left(x-y\right)\, .
\end{eqnarray}
We drop for notational convenience 
the time dependence of the bosonic field operator $\phi$, and for 
simplicity we omit the hat 
notation for the operators. 
Moreover, $m$ is the atomic mass and the interaction strength $\lambda$, proportional 
to the scattering length~\cite{Pethick20081}, 
could be taken as time-dependent. We also set $\hbar=1$.

In the Bogoliubov approximation \cite{Pethick20081,ZAGREBNOV2001291}, 
the field operator is split into two contributions: 
a classical mean-field and a quantum fluctuation field 
(with vanishing expectation value)
\begin{eqnarray}
  \qquad \qquad 
  \phi
  \left(x\right)&=&\left\langle \phi\left(x\right)\right\rangle +\delta
  \phi
  \left(x\right)\, , \label{Mean-Field splitting}\\
  \left[\delta
    \phi
    \left(x\right),\delta
    \phi
    ^{\dagger}\left(y\right)\right]&=&\delta\left(x-y\right)\, .
\end{eqnarray}
The exact equations for the dynamics of these objects could be obtained from the full Eq.~\eref{Full dynamics}, but two approximations ought to be considered
\begin{eqnarray}
\fl
i\partial_{t}\left\langle \phi\right\rangle &=& -\frac{\nabla^{2}}{2m}\left\langle \phi\right\rangle +\lambda\left\langle \phi^{\dagger}\phi\phi\right\rangle \approx-\frac{\nabla^{2}}{2m}\left\langle \phi\right\rangle +\lambda\overline{\left\langle \phi\right\rangle }\left\langle \phi\right\rangle \left\langle \phi\right\rangle \label{Gross-Pitaevskii}\, ,\\
\fl
i\partial_{t}\delta\phi &=& -\frac{\nabla^{2}}{2m}\delta\phi+\lambda\left(\phi^{\dagger}\phi\phi-\left\langle \phi^{\dagger}\phi\phi\right\rangle \right)\approx-\frac{\nabla^{2}}{2m}\left\langle \phi\right\rangle +2\lambda\overline{\left\langle \phi\right\rangle }\left\langle \phi\right\rangle \delta\phi+\lambda\left\langle \phi\right\rangle ^{2}\delta\phi^{\dagger}\label{Bogoliubov-de Gennes}
\end{eqnarray}
(the bar denoting complex conjugation). 
The first equation is
the Gross-Pitaevskii equation, 
and we refer to the second
as the Bogoliubov-de Gennes (operator) equation. 
The mean-field term represents the condensate wavefunction, 
and the approximation in Eq.~\eref{Gross-Pitaevskii} is to 
remove from the evolution the backreaction of the
fluctuation on the condensate. The second approximation is to drop all the non-linear terms (of order
higher than $\delta\phi$) from Eq.~\eref{Bogoliubov-de Gennes} which then should be diagonalised to solve
the time evolution.

A setup for analogue gravity with this kind of systems is obtained by
studying the quantum fluctuations of the phase of the condensate. 
To do so, it is better to rewrite the mean-field and the fluctuations
in terms of number density and phase, in the so-called Madelung representation:
\begin{eqnarray}
\qquad \qquad \left\langle \phi\right\rangle &= \rho_{0}^{1/2}e^{i\theta_{0}}\, ,\\
\qquad \qquad \delta\phi &= \rho_{0}^{1/2}e^{i\theta_{0}}\left(\frac{\rho_{1}}{2\rho_{0}}+i\theta_{1}\right)\, ,\\
\left[\theta_{1}\left(x\right),\rho_{1}\left(y\right)\right]&= -i\delta\left(x-y\right)\, . \label{canonical_comm_rel}
\end{eqnarray}
From the Gross-Pitaevskii equation we obtain two equations for the real classical fields $\theta_{0}$ and $\rho_{0}$:
\begin{eqnarray}
\partial_{t}\rho_{0}&=& -\frac{1}{m}\nabla\left(\rho_{0}\nabla\theta_{0}\right)\, ,\label{Continuity equation}\\
\partial_{t}\theta_{0}&=& -\lambda\rho_{0}+\frac{1}{2m}\rho_{0}^{-1/2}\left(\nabla^{2}\rho_{0}^{1/2}\right)-\frac{1}{2m}\left(\nabla\theta_{0}\right)\left(\nabla\theta_{0}\right)\, . \label{Bernoulli equation}
\end{eqnarray}
These are the quantum Euler equations for the superfluid.
Eq.~\eref{Continuity equation}
can be easily interpreted as a continuity equation for the density of the condensate, while 
Eq.~\eref{Bernoulli equation} is the Bernoulli equation for the phase 
of the superfluid, which generates the potential flow: the superfluid has
velocity $\left(\nabla\theta_{0}\right)/m$.
From the Bogoliubov-de Gennes equation we obtain two equations for
the real quantum fields $\theta_{1}$ and $\rho_{1}$
\begin{eqnarray}
\fl
\partial_{t}\rho_{1}&&= -\frac{1}{m}\nabla\left(\rho_{1}\nabla\theta_{0}+\rho_{0}\nabla\theta_{1}\right)\, ,\\
\fl
\partial_{t}\theta_{1}&&= -\left(\lambda\rho_{0}+\frac{1}{4m}\nabla\left(\rho_{0}^{-1}\left(\nabla\rho_{0}\right)\right)\right)\frac{\rho_{1}}{\rho_{0}}+\frac{1}{4m}\nabla\left(\rho_{0}^{-1}\left(\nabla\rho_{1}\right)\right)-\frac{1}{m}\left(\nabla\theta_{0}\right)\left(\nabla\theta_{1}\right)\, . \label{dynamical_theta_1}
\end{eqnarray}
If in Eq.~\eref{dynamical_theta_1} the ``quantum pressure'' term
$\nabla\left(\rho_{0}^{-1}\left(\nabla\rho_{1}\right)\right)/4m$
is negligible, as usually assumed, by substitution we obtain
\begin{eqnarray}
\fl
\rho_{1}&=-\frac{\rho_{0}}{\lambda\rho_{0}+\frac{1}{4m}\nabla\left(\rho_{0}^{-1}\left(\nabla\rho_{0}\right)\right)}\left(\left(\partial_{t}\theta_{1}\right)+\frac{1}{m}\left(\nabla\theta_{0}\right)\left(\nabla\theta_{1}\right)\right) \, ,\\
\fl
0&=  \partial_{t}\left(\frac{\rho_{0}}{\lambda\rho_{0}+\frac{1}{4m}\nabla\left(\rho_{0}^{-1}\left(\nabla\rho_{0}\right)\right)}\left(\left(\partial_{t}\theta_{1}\right)+\frac{1}{m}\left(\nabla\theta_{0}\right)\left(\nabla\theta_{1}\right)\right)\right)
\nonumber \\*
\fl
 & \phantom{=} +\nabla\left(\frac{\rho_{0}}{\lambda\rho_{0}+\frac{1}{4m}\nabla\left(\rho_{0}^{-1}\left(\nabla\rho_{0}\right)\right)}\frac{1}{m}\left(\nabla\theta_{0}\right)\left(\left(\partial_{t}\theta_{1}\right)+\frac{1}{m}\left(\nabla\theta_{0}\right)\left(\nabla\theta_{1}\right)\right)-\frac{\rho_{0}}{m}\left(\nabla\theta_{1}\right)\right)\label{Analogue Klein-Gordon}
\end{eqnarray}
with Eq.~\eref{Analogue Klein-Gordon} 
being a Klein-Gordon equation
for the field $\theta_{1}$. Eq.~\eref{Analogue Klein-Gordon} can be written in the form 
\begin{equation}
\frac{1}{\sqrt{-g}}\partial_{\mu}\left(\sqrt{-g}g^{\mu\nu}\partial_{\nu}\theta_{1}\right)=  0\, ,
\end{equation}
where the prefactor $1/\sqrt{-g}$ has been multiplied away.
This equation describes an analogue system for a scalar field in curved spacetime
as the quantum field $\theta_{1}$ propagates on
a curved geometry with a metric given by
\begin{eqnarray}
\quad \widetilde{\lambda}&&= \lambda+\frac{1}{4m}\rho_{0}^{-1}\nabla\left(\rho_{0}^{-1}\left(\nabla\rho_{0}\right)\right)\, ,\\
\quad v_{i}&&= \frac{1}{m}\left(\nabla\theta_{0}\right)_{i}\, ,\\
\sqrt{-g}&&= \sqrt{\frac{\rho_{0}^{3}}{m^{3}\widetilde{\lambda}}}\, ,\\
\quad g_{tt}&&= -\sqrt{\frac{\rho_{0}}{m\widetilde{\lambda}}}\left(\frac{\widetilde{\lambda}\rho_{0}}{m}-v^{2}\right)\, ,\\
\quad g_{ij}&&= \sqrt{\frac{\rho_{0}}{m\widetilde{\lambda}}}\delta_{ij}\, ,\\
\quad g_{ti}&&= -\sqrt{\frac{\rho_{0}}{m\widetilde{\lambda}}}v_{i}\, .
\end{eqnarray}

If  the condensate is homogeneous the superfluid velocity
vanishes, the coupling is homogeneous in space ($\widetilde{\lambda}=\lambda$),
the number density $\rho_0$ is constant in time and the only 
relevant behaviour of the condensate wavefunction is in the time-dependent phase $\theta_0$. 
Furthermore, in this case there is also no need to neglect the quantum pressure term in 
Eq.~\eref{dynamical_theta_1}, as it will be handled
easily after Fourier-transforming and it will simply introduce a modified
dispersion relation --- directly derived from the Bogoliubov
spectrum.

If the condensate has an initial uniform number density but is not
homogeneous --- meaning that the initial phase depends on
the position --- the evolution will introduce inhomogeneities
in the density $\rho_{0}$ as described by the continuity equation
Eq.~\eref{Continuity equation}, and the initial configuration will
be deformed in time. However,
as long as $\nabla^{2}\theta_{0}$ is small, also the
variations of $\rho_{0}$ are small as well: while there is not a non-trivial
stationary analogue metric, the scale of the inhomogeneities will
define a timescale for which one could safely assume stationarity.
Furthermore, the presence of an external potential $V_{ext}(x)$
in the Hamiltonian via a term of the form  
$\int\mathrm{d}x V_{ext}\left( x \right)\phi^{\dagger}\left(x\right)\phi\left( x \right)$ would
play a role in the dynamical equation for $\theta_{0}$,
leaving invariant those for $\rho_{0}$, $\rho_{1}$
and $\theta_{1}$.

\section{Time-dependent natural orbitals}\label{sec:orbitas}
The mean-field approximation presented in the previous Section 
is a solid and consistent formulation
for studying weakly interacting BEC~\cite{Leggett}. 
It requires, however, that the quantum state has peculiar features which need to be 
taken into account. In analogue gravity these assumptions are tacitly 
considered, but since they play a crucial role for our treatment we 
present a discussion of them in some detail 
to lay down the ground and the formalism 
in view of next Sections.

As it is well known~\cite{Leggett}, the mean-field approximation,  
consisting in substituting the operator $\phi(x)$ with its expectation value 
 $\left\langle \phi\left(x\right)\right\rangle$, is strictly valid when the state 
considered is coherent,
meaning it is an eigenstate of the atomic field operator $\phi$:
\begin{equation}
\phi\left(x\right)\left|coh\right\rangle =   \left\langle \phi\left(x\right)\right\rangle  \left|coh\right \rangle\, .
\end{equation}
For states satisfying this equation, the Gross-Pitaevskii equation
\eref{Gross-Pitaevskii} is exact (while Eq.~\eref{Bogoliubov-de Gennes}
is still a linearised approximation). 
We remind that the coherent states $\left|coh\right\rangle $
are not eigenstates of the number operator,
but they are rather quantum superpositions of states with different number of
atoms. This is necessary because $\phi$ is an operator that ---
in the non relativistic limit --- destroys a particle. 
We also observe that the notion of coherent state is valid instantaneously,
but it may be in general
not preserved along the evolution in presence of an interaction. 

The redefinition of the field operator as 
in Eq.~\eref{Mean-Field splitting}
provides a description where the physical degrees of
freedom are concealed: the new degrees of freedom are not the excited
atoms, but the quantized fluctuations over a coherent state. 
Formally this is a simple and totally legit redefinition, but for
our discussion we stress that
the quanta created by the operator $\delta\phi$ do not have a direct interpretation
as atoms.

Given the above discussion, it is useful to remember 
that coherent states are not the only states to express the condensation, 
\emph{i.e.} the fact that a macroscopic number of particles occupies 
the same state. 
As it is stated in the Penrose-Onsager criterion for off-diagonal 
long-range order~\cite{PhysRev.104.576,pitaevskii2003bose}, the condensation phenomenon 
is best defined considering the properties
of the $2$-point correlation functions. 

The $2$-point correlation function is the expectation value on the 
quantum state of an operator
composed of the creation of a particle in a position $x$ after the
destruction of a particle in a different position $y$: 
$\left\langle \phi^{\dagger}\left(x\right)\phi\left(y\right)\right\rangle $.
Since by definition the $2$-point correlation function is Hermitian, $\overline{\left\langle \phi^{\dagger}\left(y\right)\phi\left(x\right)\right\rangle }= \left\langle \phi^{\dagger}\left(x\right)\phi\left(y\right)\right\rangle 
$, it can always be diagonalised as
\begin{equation}
  \left\langle \phi^{\dagger}\left(x\right)\phi\left(y\right)\right\rangle =\sum_{I}\left\langle N_{I}\right\rangle \overline{f_{I}}\left(x\right)f_{I}\left(y\right)\label{2-point correlation function},
\end{equation}
with
\begin{equation}
\int\mathrm{d}x\overline{f_{I}}\left(x\right)f_{J}\left(x\right)= \delta_{IJ}\, .
\end{equation}
The orthonormal functions $f_{I}$, eigenfunctions of the $2$-point
correlation function, are known as the natural orbitals, and
define a complete basis for the $1$-particle Hilbert space. In the 
case of a time-dependent Hamiltonian (or during the dynamics)
they are in turn time-dependent. As 
for the field operator, to simplify the notation we are not going to always 
write explicitly the time dependence of the $f_{I}$. 

The eigenvalues
$\left\langle N_{I}\right\rangle $ are the occupation numbers of
these wavefunctions. The sum of these eigenvalues gives the total
number of particles in the state (or the mean value, in the case of
superposition of quantum states with different number of particles):
\begin{equation}
\left\langle N\right\rangle=  \sum_{I}\left\langle N_{I}\right\rangle\, . 
\end{equation}
The time-dependent orbitals define creation and destruction operators
and consequently the relative number operators (having as expectation
values the eigenvalues of the $2$-point correlation function):
\begin{eqnarray}
\qquad a_{I}&= \int\mathrm{d}x\overline{f_{I}}\left(x\right)\phi\left(x\right)\, , \\
\Bigl[a_{I},a_{J}^{\dagger}\Bigr]&= \delta_{IJ} \label{comm_aadagg}\, ,\\
\Bigl[a_{I},a_{J}\Bigr]&= 0 \label{comm_aa}\, ,\\
\qquad N_{I}&= a_{I}^{\dagger}a_{I}\, .
\end{eqnarray}
The state is called ``condensate''~\cite{PhysRev.104.576}
when one of these occupation numbers 
is macroscopic (comparable with the total number of particles)
and the others are small when compared to it. 

In the weakly interacting limit,
the condensed fraction $\left\langle N_{0}\right\rangle /\left\langle N\right\rangle$
is approximately equal to $1$, and the depletion factor 
$\sum_{I\neq0}\left\langle N_{I}\right\rangle /\left\langle N\right\rangle$
is negligible.
This requirement is satisfied by coherent states which define perfect
condensates since the $2$-point correlation functions is exactly
a product of the mean-field and its conjugate:
\begin{equation}
\left\langle coh\right|\phi^{\dagger}\left(x\right)\phi\left(y\right)\left|coh\right\rangle = \overline{\left\langle \phi\left(x\right)\right\rangle }\left\langle \phi\left(y\right)\right\rangle \label{2ptCF_coh}\, ,
\end{equation}
with
\begin{eqnarray}
f_{0}\left(x\right)&= \left\langle N_{0}\right\rangle ^{-1/2}\left\langle \phi\left(x\right)\right\rangle \, , \\
\left\langle N_{0}\right\rangle &= \int\mathrm{d}y\overline{\left\langle \phi\left(y\right)\right\rangle }\left\langle \phi\left(y\right)\right\rangle \, ,\\
\left\langle N_{I\neq0}\right\rangle &= 0 \, .
\end{eqnarray}
Therefore, in this case, the set of time-dependent orbitals is given
by the proper normalisation of the mean-field function with a completion
that is the basis for the subspace of the Hilbert space orthogonal
to the mean-field. The latter set can be redefined arbitrarily, as
the only non-vanishing eigenvalue of the $2$-point correlation function
is the one relative to mean-field function. 
The fact that there is a non-vanishing macroscopic eigenvalue implies that 
there is total condensation, \emph{i.e.} $\left\langle N_{0}\right\rangle /\left\langle N\right\rangle=1$.

\subsection{Time-dependent orbitals formalism}

It is important to understand how we can study the condensate state even 
if we are not considering coherent states 
and how the description is related to the mean-field approximation. 
In this framework, we shall see that the mean-field 
approximation is not a strictly 
necessary theoretical requirement for analogue gravity.

With respect to the basis of time-dependent orbitals and their creation 
and destruction operators, we can introduce a new expression for 
the atomic field operator, projecting it on the sectors of the Hilbert space 
as
\begin{eqnarray}
\phi\left(x\right) & =\phi_{0}\left(x\right)+\phi_{1}\left(x\right)\nonumber\\
 & =f_{0}\left(x\right)a_{0}+\sum_{I}f_{I}\left(x\right)a_{I}\nonumber\\
 & =f_{0}\left(x\right)\left(\int\mathrm{d}y\overline{f_{0}}\left(y\right)\phi\left(y\right)\right)+\sum_{I\neq0}f_{I}\left(x\right)\left(\int\mathrm{d}y\overline{f_{I}}\left(y\right)\phi\left(y\right)\right)\, . 
\label{form}
\end{eqnarray}

The two parts of the atomic field operator so defined are related
to the previous mean-field $\left\langle \phi\right\rangle $ and
quantum fluctuation $\delta\phi$ expressions given in 
Section~\ref{sec:AG-BEC}. 
One has now the commutation relation
\begin{equation}
\left[\phi_{0}\left(x\right),\phi_{0}^{\dagger}\left(y\right)\right]= f_{0}\left(x\right)\overline{f_{0}}\left(y\right)= O\left( V^{-1}\right)\, ,  
\label{phi_0-commutator}
\end{equation}
where $V$ denotes the volume of the system. 
Observe that while the commutator~\eref{phi_0-commutator}  
does not vanish identically, it is negligible in the limit of large $V$.

In the formalism~\eref{form} the condensed part of the field 
is described by the operator $\phi_{0}$
and by the orbital producing it through projection, the $1$-particle
wave function $f_0$. The dynamics of the function $f_{0}$, 
the $1$-particle wave function
that best describes the collective behaviour of the condensate, 
can be extracted by using the relations 
\begin{eqnarray}
\left\langle \phi^{\dagger}\left(x\right)\phi\left(y\right)\right\rangle &= \sum_{I}\left\langle N_{I}\right\rangle \overline{f_{I}}\left(x\right)f_{I}\left(y\right)\, ,\\
\left\langle \left[a_{K}^{\dagger}a_{J},H\right]\right\rangle &= i\partial_{t}\left\langle N_{J}\right\rangle \delta_{JK}+i\left(\left\langle N_{K}\right\rangle -\left\langle N_{J}\right\rangle \right)\left(\int\mathrm{d}x\overline{f_{J}}\left(x\right)\dot{f_{K}}\left(x\right)\right)\, 
\end{eqnarray}
and the evolution of the condensate 1-particle wave function
\begin{eqnarray}
\fl
i\partial_{t}f_{0}\left(x\right)&= \left(\int\mathrm{d}y\overline{f_{0}}\left(y\right)\left(i\partial_{t}f_{0}\left(y\right)\right)\right)f_{0}\left(x\right)+\sum_{I\neq0}\left(\int\mathrm{d}y\overline{f_{0}}\left(y\right)\left(i\partial_{t}f_{I}\left(y\right)\right)\right)f_{I}\left(x\right)\nonumber 
\\
\fl
&= \left(\int\mathrm{d}y\overline{f_{0}}\left(y\right)\left(i\partial_{t}f_{0}\left(y\right)\right)\right)f_{0}\left(x\right) +\sum_{I\neq0}\frac{1}{\left\langle N_{0}\right\rangle -\left\langle N_{I}\right\rangle }\left\langle \left[a_{0}^{\dagger}a_{I},H\right]\right\rangle f_{I}\left(x\right) \nonumber\\
\fl
&=-\frac{i}{2}\frac{\partial_{t}\left\langle N_{0}\right\rangle }{\left\langle N_{0}\right\rangle }f_{0}\left(x\right)+ \left(-\frac{\nabla^{2}}{2m}f_{0}\left(x\right)\right)+\frac{1}{\left\langle N_{0}\right\rangle }\left\langle a_{0}^{\dagger}\left[\phi\left(x\right),V\right]\right\rangle + \nonumber \\
\fl
&\qquad+\sum_{I\neq0}\frac{\left\langle N_{I}\right\rangle \left\langle a_{0}^{\dagger}\left[a_{I},V\right]\right\rangle +\left\langle N_{0}\right\rangle \left\langle \left[a_{0}^{\dagger},V\right]a_{I}\right\rangle }{\left\langle N_{0}\right\rangle \left(\left\langle N_{0}\right\rangle -\left\langle N_{I}\right\rangle \right)}f_{I}\left(x\right)
\label{time derivative f_0 final}
\end{eqnarray}
(we assumed at any time 
$\left\langle N_{0}\right\rangle \neq\left\langle N_{I\neq0}\right\rangle $). 
The above equation is valid for a condensate when the dynamics is driven by a Hamiltonian 
operator composed of a kinetic term and a generic potential $V$, but we are interested in the case 
of Eq.~\eref{MB Hamiltonian}. 
Furthermore, $f_{0}\left(x\right)$ can be
redefined through an overall phase transformation, 
$f_{0}\left(x\right)\rightarrow  e^{i\Theta}f_{0}\left(x\right)$ with any arbitrary time-dependent real function $\Theta$. 
We have chosen the overall phase to satisfy
the final expression Eq.~\eref{time derivative f_0 final}, as it is
the easiest to compare with the Gross-Pitaevskii equation~\eref{Gross-Pitaevskii}.

\subsection{Connection with the Gross-Pitaevskii equation}
\label{connection}

In this Section we discuss the relation between 
the function $f_{0}$ 
--- the eigenfunction of the $2$-point correlation function with 
macroscopic eigenvalue --- and the solution of the Gross-Pitaevskii equation, approximating the mean-field function for quasi-coherent states.
In particular, we aim at comparing the equations describing their dynamics, 
detailing under which approximations they show the same behaviour. This 
discussion provides a preliminary technical basis for the study of the effect 
of the quantum correlations between the background condensate and the 
quasi-particles, which are present when the quantum nature of the 
condensate field operator is retained and it is not just approximated 
by a number, as done when performing the standard
Bogoliubov approximation. We refer 
to~\cite{ZAGREBNOV2001291} for a review on 
the Bogoliubov approximation in weakly imperfect Bose gases and 
to~\cite{Lieb2006} for a presentation of rigorous results 
on the condensation properties of dilute bosons.

The Gross-Pitaveskii equation Eq.~\eref{Gross-Pitaevskii} 
is an approximated equation for the mean-field dynamics. 
It holds when the backreaction
of the fluctuations $\delta\phi$ on the condensate --- described 
by a coherent state --- is negligible. This equation
includes a notion of number conservation, meaning that the approximation 
of the interaction term implies that the spatial integral of the squared
norm of the solution of the equation is conserved
\begin{eqnarray}
\quad \left\langle \phi^{\dagger}\left(x\right)\phi\left(x\right)\phi\left(x\right)\right\rangle &\approx& \overline{\left\langle \phi\left(x\right)\right\rangle }\left\langle \phi\left(x\right)\right\rangle \left\langle \phi\left(x\right)\right\rangle \\
&\Downarrow & \nonumber \\
i\partial_{t}\int\mathrm{d}x\overline{\left\langle \phi\left(x\right)\right\rangle }\left\langle \phi\left(x\right)\right\rangle &
=0\, .
\end{eqnarray}
This depends on the fact that only the leading term of the interaction is included in the equation. Therefore we can 
compare the Gross-Pitaevskii
equation for the mean-field with the equation for 
$\left\langle N_{0}\right\rangle ^{1/2}f_{0}\left(x\right)$ approximated to
leading order, \emph{i.e.} $\left\langle \phi\left(x\right)\right\rangle$
should be compared to the function $f_{0}\left(x\right)$ 
under the approximation that there is no depletion from the condensate. 
If we consider the approximations
\begin{eqnarray}
\left\langle a_{0}^{\dagger}\left[\phi\left(x\right),V\right]\right\rangle \approx & \lambda\left\langle N_{0}\right\rangle ^{2}\overline{f_{0}}\left(x\right)f_{0}\left(x\right)f_{0}\left(x\right) \, ,\label{approx1}
\end{eqnarray}
\begin{eqnarray}
i\partial_{t}\left\langle N_{0}\right\rangle =\left\langle \left[a_{0}^{\dagger}a_{0},V\right]\right\rangle \approx & 0 \, ,\label{approx2}
\end{eqnarray}
\begin{eqnarray}
\sum_{I\neq0}\frac{\left\langle N_{I}\right\rangle \left\langle a_{0}^{\dagger}\left[a_{I},V\right]\right\rangle +\left\langle N_{0}\right\rangle \left\langle \left[a_{0}^{\dagger},V\right]a_{I}\right\rangle }{\left\langle N_{0}\right\rangle -\left\langle N_{I}\right\rangle }f_{I}\left(x\right)\approx & 0 \, ,\label{approx3}
\end{eqnarray}
we obtain that $\left\langle N_{0}\right\rangle^{1/2}f_{0}\left(x\right)$ 
satisfies the Gross-Pitaevskii equation.

The approximation in Eq.~\eref{approx1}  is easily justified, 
since we are retaining only the leading order of the expectation value 
$\left\langle a_{0}^{\dagger}\left[\phi,V\right]\right\rangle $,
and neglecting the others which depend  on the operators $\phi_{1}$ and $\phi_{1}^{\dagger}$ and are of order smaller than $\left\langle N_{0}\right\rangle ^{2}$.
The second equation Eq.~\eref{approx2} is derived from the previous one as
a direct consequence, since the depletion of $N_{0}$ comes from the subleading terms $\left\langle a_{0}^{\dagger}\phi_{1}^{\dagger}\phi_{1}a_{0}\right\rangle $
and $\left\langle a_{0}^{\dagger}a_{0}^{\dagger}\phi_{1}\phi_{1}\right\rangle $.
The first of these two terms is of order $\left\langle N_{0}\right\rangle $,
having its main contributions from separated diagrams --- $\left\langle a_{0}^{\dagger}\phi_{1}^{\dagger}\phi_{1}a_{0}\right\rangle \approx\left\langle N_{0}\right\rangle \left\langle \phi_{1}^{\dagger}\phi_{1}\right\rangle $ --- and the second is of the same order due to the dynamics. 
The other terms are even more suppressed, as can be argued considering that they contain an odd number of operators $\phi_{1}$. Taking their time derivatives, we observe that they arise from 
the second order in the interaction, making these terms negligible in the regime of weak interaction. The terms 
$\left\langle a_{0}^{\dagger}a_{0}^{\dagger}a_{0}\phi_{1}\right\rangle $
are also sub-leading with respect to those producing the depletion, 
since the contributions coming from separated diagrams vanish by definition,
and the remaining fully connected diagrams describe the correlation 
between small operators, acquiring relevance only while the many-body quantum
state is mixed by the depletion of the condensate:
\begin{eqnarray}
\quad \qquad \left\langle a_{0}^{\dagger}a_{0}^{\dagger}a_{0}\phi_{1}\right\rangle  & =\left\langle a_{0}^{\dagger}\phi_{1}\left(N_{0}-\left\langle N_{0}\right\rangle \right)\right\rangle \, .
\end{eqnarray}
Using the same arguments we can assume the approximation in Eq.~\eref{approx3} to hold, as the denominator of order $\left\langle N_{0}\right\rangle $
is sufficient to suppress the terms in the numerator, which are negligible with respect to the leading term in Eq.~\eref{approx1}. 

The leading terms in Eq.~\eref{approx3} do not affect the depletion of $N_{0}$, but they may be of the same order. They depend on the expectation value
\begin{eqnarray}
\left\langle a_{0}^{\dagger}\left[a_{I},V\right]\right\rangle &\approx\lambda\left(\int\mathrm{d}x\overline{f_{I}}\left(x\right)\overline{f_{0}}\left(x\right)f_{0}\left(x\right)f_{0}\left(x\right)\right)\left\langle N_{0}\right\rangle ^{2}\, .
\end{eqnarray}
 Therefore these mixed terms --- with two ladder operators relative one to the condensed state and one
  to an excited state --- are completely negligible when the integral $\int\overline{f_{I}}\overline{f_{0}}f_{0}f_{0}$ 
  is sufficiently small. This happens when the condensed 1-particle state is approximately $f_{0}\approx V^{-1/2}e^{i\theta_{0}}$, \emph{i.e.} when
   the atom number density of the condensate is approximately homogeneous.

Moreover, in many cases of interest, it often holds 
that the terms in the LHS of Eq.~\eref{approx3} vanish identically: if the quantum state 
is an eigenstate of a conserved charge --- \emph{e.g.} total momentum or total angular momentum --- 
the orbitals must be labeled with a specific value of charge. The relative ladder operators act by 
adding or removing from the state such charge, and for any expectation value not to vanish the
charges must cancel out. In the case of homogeneity of the condensate 
and translational invariance of the Hamiltonian, this statement regards
the conservation of momentum. 
In particular, if the state is invariant under translations, we have
\begin{eqnarray}
\left\langle a_{k_{1}}^{\dagger}a_{k_{2}}^{\dagger}a_{k_{3}}a_{k_{4}}\right\rangle &= \delta_{k_{1}+k_{2},k_{3}+k_{4}}\left\langle a_{k_{1}}^{\dagger}a_{k_{2}}^{\dagger}a_{k_{3}}a_{k_{4}}\right\rangle \, ,\\
\quad \left\langle a_{0}^{\dagger}a_{0}^{\dagger}a_{0}a_{k}\right\rangle &= 0 
\end{eqnarray}
(for $k \neq 0$). 

In conclusion, we obtain that for a condensate with a quasi-homogeneous density a good approximation 
for the dynamics of the function $\left\langle N_{0}\right\rangle ^{1/2}f_{0}$, 
the rescaled $1$-particle wavefunction macroscopically 
occupied by the condensate, is provided by
\begin{eqnarray}
\fl
i\partial_{t}\left(\left\langle N_{0}\right\rangle ^{1/2}f_{0}\left(x\right)\right) & =-\frac{\nabla^{2}}{2m}\left(\left\langle N_{0}\right\rangle ^{1/2}f_{0}\left(x\right)\right)+\lambda\left\langle N_{0}\right\rangle ^{3/2}\overline{f_{0}}\left(x\right)f_{0}\left(x\right)f_{0}\left(x\right)+ \nonumber \\
\fl
& \phantom{=}+\lambda\left\langle N_{0}\right\rangle ^{-1}\left(\left\langle a_{0}^{\dagger}\phi_{1}\left(x\right)\phi_{1}^{\dagger}\left(x\right)a_{0}\right\rangle+\left\langle a_{0}^{\dagger}\phi_{1}^{\dagger}\left(x\right)\phi_{1}\left(x\right)a_{0}\right\rangle \right) \left(\left\langle N_{0}\right\rangle ^{1/2}f_{0}\left(x\right)\right)+\nonumber \\
\fl
& \phantom{=}+\lambda\left\langle N_{0}\right\rangle ^{-1}\left\langle a_{0}^{\dagger}\phi_{1}\left(x\right)a_{0}^{\dagger}\phi_{1}\left(x\right)\right\rangle \left(\left\langle N_{0}\right\rangle ^{1/2}\overline{f_{0}}\left(x\right)\right)+O\left(\lambda\left\langle N_{0}\right\rangle ^{1/2}V^{-3/2}\right) \, . \nonumber \\ \* \fl \label{dynamics-rescaled-f_0}
\end{eqnarray}
This equation is equivalent to the Gross-Pitaveskii equation Eq.~\eref{Gross-Pitaevskii} when we consider only the leading terms, \emph{i.e.}~the first line of Eq.~\eref{dynamics-rescaled-f_0}. If we also consider the remaining
lines of the Eq.~\eref{dynamics-rescaled-f_0}, \emph{i.e.} if we include the effect of the depletion, we obtain an equation which should be compared to the equation for the mean-field function up to the terms quadratic in the operators $\delta \phi$.
The two equations are analogous when making the identification
\begin{eqnarray}
\quad \left\langle N_{0}\right\rangle ^{1/2}f_{0}&\sim \left\langle \phi\right\rangle \, ,\\
\left\langle \phi_{0}^{\dagger}\phi_{1}^{\dagger}\phi_{1}\phi_{0}\right\rangle &\sim \overline{\left\langle \phi\right\rangle }\left\langle \phi\right\rangle \left\langle \delta\phi^{\dagger}\delta\phi\right\rangle \, ,\\
\left\langle \phi_{0}^{\dagger}\phi_{0}^{\dagger}\phi_{1}\phi_{1}\right\rangle &\sim \overline{\left\langle \phi\right\rangle }\overline{\left\langle \phi\right\rangle }\left\langle \delta\phi\delta\phi\right\rangle \, .
\end{eqnarray}
The possible ambiguities in comparing the two equations come from the arbitrariness in fixing the over-all time-dependent phases of
the functions $f_{0}$ and $\left\langle \phi\right\rangle $, and 
from the fact that the commutation relations for the operators $\phi_{1}$
and the operators $\delta\phi$ differ from each other by 
a term going as $\overline{f_{0}}f_{0}$, as seen in Eq.~\eref{phi_0-commutator}. 
This causes an apparent difference when comparing the two terms 
\begin{eqnarray}
\fl
&&\lambda\left\langle N_{0}\right\rangle ^{-1}\left(\left\langle a_{0}^{\dagger}\phi_{1}\left(x\right)\phi_{1}^{\dagger}\left(x\right)a_{0}\right\rangle +\left\langle a_{0}^{\dagger}\phi_{1}^{\dagger}\left(x\right)\phi_{1}\left(x\right)a_{0}\right\rangle \right)\left(\left\langle N_{0}\right\rangle ^{1/2}f_{0}\left(x\right)\right)\nonumber\\
&&\sim  2\lambda\left\langle \delta\phi^{\dagger}\left(x\right)\delta\phi\left(x\right)\right\rangle \left\langle \phi\left(x\right)\right\rangle \, .\fl 
\end{eqnarray}
However the difference can be reabsorbed --- manipulating the
RHS --- in a term which only affects the overall phase of the mean-field, not the superfluid velocity.

The equation for the depletion can be easily derived for the number-conserving approach and compared to the result in the Bogoliubov approach.
As we have seen, the dynamical equation for $f_{0}$ contains the 
information for the time derivative of its occupation number. 
Projecting the derivative along the function itself and taking the imaginary part, one gets
\begin{eqnarray}
\fl
\frac{1}{2}i\partial_{t}\left\langle N_{0}\right\rangle  & =\frac{1}{2}\left\langle \left[a_{0}^{\dagger}a_{0},V\right]\right\rangle \nonumber\\
 \fl
 & =i\mathrm{Im}\left(\int\mathrm{d}x\left\langle N_{0}\right\rangle ^{1/2}\overline{f_{0}}\left(x\right)i\partial_{t}\left(\left\langle N_{0}\right\rangle ^{1/2}f_{0}\left(x\right)\right)\right)\nonumber\\
 \fl
 & \approx\frac{\lambda}{2}\left(\int\mathrm{d}x\left\langle \phi_{0}^{\dagger}\left(x\right)\phi_{0}^{\dagger}\left(x\right)\phi_{1}\left(x\right)\phi_{1}\left(x\right)\right\rangle -\left\langle \phi_{1}^{\dagger}\left(x\right)\phi_{1}^{\dagger}\left(x\right)\phi_{0}\left(x\right)\phi_{0}\left(x\right)\right\rangle \right) \, .
\end{eqnarray}
We can now compare this equation to the one for the depletion in 
mean-field description
\begin{eqnarray}
\fl
\frac{1}{2}i\partial_{t}N & =\frac{1}{2}i\partial_{t}\left(\int\mathrm{d}x\overline{\left\langle \phi\left(x\right)\right\rangle }\left\langle \phi\left(x\right)\right\rangle \right)\nonumber\\
\fl
& =i\mathrm{Im}\left(\int\overline{\left\langle \phi\left(x\right)\right\rangle }i\partial_{t}\left\langle \phi\left(x\right)\right\rangle \right)\nonumber\\
\fl
& \approx\frac{\lambda}{2}\left(\int\overline{\left\langle \phi\left(x\right)\right\rangle }\overline{\left\langle \phi\left(x\right)\right\rangle }\left\langle \delta\phi\left(x\right)\delta\phi\left(x\right)\right\rangle -\left\langle \phi\left(x\right)\right\rangle \left\langle \phi\left(x\right)\right\rangle \left\langle \delta\phi^{\dagger}\left(x\right)\delta\phi^{\dagger}\left(x\right)\right\rangle \right) \, .
\end{eqnarray}
The two expressions are consistent with each other
\begin{eqnarray}
\left\langle \phi_{0}^{\dagger}\left(x\right)\phi_{0}^{\dagger}\left(x\right)\phi_{1}\left(x\right)\phi_{1}\left(x\right)\right\rangle \sim & \overline{\left\langle \phi\left(x\right)\right\rangle }\overline{\left\langle \phi\left(x\right)\right\rangle }\left\langle \delta\phi\left(x\right)\delta\phi\left(x\right)\right\rangle .
\end{eqnarray}

For coherent states one expects to find equivalence between $\delta\phi$ and $\phi_{1}$. To do so we need 
to review the number-conserving formalism, that can provide the same description
used for analogue gravity in the general case, \emph{e.g.} when there is
a condensed state with features different from those of coherent states.

\section{Number-conserving formalism}\label{QP-BEC}

Within the mean-field framework, the splitting of the field obtained by translating the field operator $\phi$ by the mean-field function
produces the new field $\delta\phi$. This redefinition of the field also induces a corresponding one of the Fock space which to a certain extent
hides the physical atom degrees of freedom, since the field $\delta\phi$ describes the quantum fluctuations over the mean-field.

Analogue gravity is defined considering this field and its Hermitian conjugate, properly combined, to study the fluctuation of phase. The
fact that $\delta\phi$ is obtained by translation provides this field with canonical commutation relation.
The mean-field description for condensates holds for coherent states and is a good approximation for quasi-coherent states. 

When we consider states with fixed number of atoms, therefore not coherent states,
it is better to consider different operators to study the
fluctuations. One can do it following the intuition that the fluctuation represents a shift of a single atom from the condensate to the excited
fraction and viceversa. Our main reason here to proceed this way is that
we are forced to retain the quantum nature of the condensate. 
We therefore want to adopt the established formalism of number-conserving 
ladder operators (see \emph{e.g.}~\cite{Pethick20081})
to obtain a different expression for the
Bogoliubov-de Gennes equation, studying the excitations of the condensate in 
these terms. We can adapt this discussion to the time-dependent orbitals. 

An important remark is that the qualitative point of introducing the number-conserving approach 
 is conceptually separated from the fact 
that higher-order terms are anyway neglected by doing Bogoliubov 
approximations~\cite{Pethick20081,Proukakis_2008}. 
Indeed, even going beyond the latter, 
including terms with three or four quasi-particle operators, 
one has anyway the problem 
that neglecting the commutation relations for ${a}_{0}$ no quantum 
correlations between the quasi-particles and the condensate 
are still
present. Including such terms, in a growing level 
of accuracy (and complexity), the main difference would be that the true 
quasi-particles of the systems do no longer coincide with the Bogoliubov ones. 
From a practical point of view, this makes clearly a (possibly)
huge quantitative 
difference for the energy spectrum, correlations between quasi-particles, 
transport properties and observables. Nevertheless this does not touch at heart 
that the quantum nature of the condensate is not retained. 
A discussion of the terms one can include beyond Bogoliubov approximation, 
and the resulting hierarchy of approximations,  
is presented in~\cite{Proukakis_2008}. Here our point is rather 
of principle, \emph{i.e.}  
the investigation of the consequences of retaining the operator nature 
of ${a}_{0}$. Therefore we basically stick to the 
standard Bogoliubov approximation, improved via the introduction 
of number-conserving operators.

If we consider the ladder operators $a_{I}$, satisfying by definition the relations 
in Eq.~\eref{comm_aadagg}--\eref{comm_aa}, and keep as reference the 
state $a_{0}$ for the condensate, it is a straightforward procedure to define 
the number-conserving operators $\alpha_{I\neq0}$, one for each excited 
wavefunction, according to the relations 
\begin{eqnarray}
\qquad \alpha_{I} &= N_{0}^{-1/2}a_{0}^{\dagger}a_{I}\, ,\label{alphaa}\\
\Bigl[\alpha_{I},\alpha_{J}^{\dagger}\Bigr]&= \delta_{IJ} \qquad\qquad \forall I,J\neq0\, ,\\
\Bigl[\alpha_{I},\alpha_{J}\Bigr]&=0_{\phantom{IJ}}\qquad\qquad\forall I,J\neq0\, .
\end{eqnarray}
The degree relative to the condensate is absorbed into the definition,
from the hypothesis of number conservation. These relations hold for
$I,J\neq0$, and obviously there is no {number-conserving ladder
operator relative to the condensed state}. The operators $\alpha_{I}$
are not a complete set of operators to describe the whole Fock space,
but they span any subspace of given number of total atoms. To move
from one another it would be necessary to include the operator $a_{0}$. 

This restriction to a subspace of the Fock space is analogous to what
is implicitly done in the mean-field approximation, where one considers
the subspace of states which are coherent with respect to the action
of the destruction operator associated to the mean-field function.

In this setup we need to relate the excited part described by $\phi_{1}$
to the usual translated field $\delta\phi$, and obtain an equation
for its dynamics related to the Bogoliubov-de Gennes equation.
To do so we need to study the linearisation of the dynamics of the
operator $\phi_{1}$, combined with the proper operator providing
the number conservation
\begin{eqnarray}
N_{0}^{-1/2}a_{0}^{\dagger}\phi_{1} &= N_{0}^{-1/2}a_{0}^{\dagger}\left(\phi-\phi_{0}\right) \nonumber\\
&= \sum_{I\neq0}f_{I}\alpha_{I} \, .
\end{eqnarray}
As long as the approximations needed to write
a closed dynamical equation for $\phi_{1}$ 
are compatible with those approximations under which the equation
for the dynamics of $f_{0}$ resembles the Gross-Pitaevskii equation --- \emph{i.e.} as
long as the time derivative of the operators $\alpha_{I}$ can be
written as a combination of
the $\alpha_{I}$ themselves ---
we can expect to have a setup for analogue gravity.

Therefore we consider the order of magnitude of the various contributions
to the time derivative of $\left(N_{0}^{-1/2}a_{0}^{\dagger}\phi_{1}\right)$.
We have already discussed the time evolution of the function $f_{0}$: from
the latter it depends the evolution of the operator $a_{0}$, since it is the
projection along $f_{0}$ of the full field operator $\phi$.
At first we observe that the variation in time of $N_{0}$ must be
of smaller order, both for the definition of condensation
and because of the approximations considered in the previous section
\begin{eqnarray}
\fl
i\partial_{t}N_{0} &=  i\partial_{t}\left(\int\mathrm{d}x\mathrm{d}yf_{0}\left(x\right)\overline{f_{0}}\left(y\right)\phi^{\dagger}\left(x\right)\phi\left(y\right)\right)\nonumber\\
\fl
 &= \left[a_{0}^{\dagger}a_{0},V\right]+\sum_{I\neq0}\frac{1}{\left\langle N_{0}\right\rangle -\left\langle N_{I}\right\rangle }\left(\left\langle \left[a_{0}^{\dagger}a_{I},V\right]\right\rangle a_{I}^{\dagger}a_{0}+\left\langle \left[a_{I}^{\dagger}a_{0},V\right]\right\rangle a_{0}^{\dagger}a_{I}\right)\, .
\end{eqnarray}
The second term is negligible
due to
the prefactor $\left\langle \left[a_{I}^{\dagger}a_{0},V\right]\right\rangle $.
Moreover,
since $\left\langle a_{I}^{\dagger}a_{0}\right\rangle$ is vanishing,  
the dominant term is the first, which
has contributions of at most the order of the depletion factor.

The same can be argued for both the operators $a_{0}$ and $N_{0}^{-1/2}$:
their derivative do not provide leading terms when we consider the
derivative of the composite operator
$\left(N_{0}^{-1/2}a_{0}^{\dagger}\phi_{1}\right)$,
only the derivative of the last operator $\phi_{1}$ being relevant.
At leading order, we have 
\begin{eqnarray}
i\partial_{t}\left(N_{0}^{-1/2}a_{0}^{\dagger}\phi_{1}\right)\approx & i\left(N_{0}^{-1/2}a_{0}^{\dagger}\right)\left(\partial_{t}\phi_{1}\right)\, .
\end{eqnarray}
We therefore have to analyse the properties of $\partial_{t}\phi$,
considering the expectation values between the orthogonal components
$\phi_{0}$ and $\phi_{1}$ and their time derivatives:
\begin{eqnarray}
\fl
\left\langle \phi_{0}^{\dagger}\left(y\right)\left(i\partial_{t}\phi_{0}\left(x\right)\right)\right\rangle  & =\left(\left\langle N_{0}\right\rangle ^{1/2}\overline{f_{0}}\left(y\right)\right)i\partial_{t}\left(\left\langle N_{0}\right\rangle ^{1/2}f_{0}\left(x\right)\right) \, ,\\
\fl
\left\langle \phi_{0}^{\dagger}\left(y\right)\left(i\partial_{t}\phi_{1}\left(x\right)\right)\right\rangle  & =-\sum_{I\neq0}\overline{f_{0}}\left(y\right)f_{I}\left(x\right)\frac{\left\langle N_{I}\right\rangle \left\langle a_{0}^{\dagger}\left[a_{I},V\right]\right\rangle +\left\langle N_{0}\right\rangle \left\langle \left[a_{0}^{\dagger},V\right]a_{I}\right\rangle }{\left\langle N_{0}\right\rangle -\left\langle N_{I}\right\rangle }
\nonumber\\
\fl
&=-\left\langle \left(i\partial_{t}\phi_{0}^{\dagger}\left(y\right)\right)\phi_{1}\left(x\right)\right\rangle \, ,\\
\fl
\left\langle \phi_{1}^{\dagger}\left(y\right)\left(i\partial_{t}\phi_{1}\left(x\right)\right)\right\rangle  & =\left\langle \phi_{1}^{\dagger}\left(y\right)\left(-\frac{\nabla^{2}}{2m}\phi_{1}\left(x\right)\right)\right\rangle +\left\langle \phi_{1}^{\dagger}\left(y\right)\left[\phi_{1}\left(x\right),V\right]\right\rangle
\nonumber \\
\fl
&\phantom{=} -\sum_{I\neq0}\overline{f_{I}}\left(y\right)f_{0}\left(x\right)\frac{\left\langle N_{I}\right\rangle \left\langle \left[a_{I}^{\dagger}a_{0},V\right]\right\rangle }{\left\langle N_{0}\right\rangle -\left\langle N_{I}\right\rangle }\, .
\end{eqnarray}
The first equation shows that
the function $\left\langle N_{0}\right\rangle ^{1/2}f_{0}\left(x\right)$
assumes the same role of the solution of Gross-Pitaevskii
equation in the mean-field description. As long as the expectation
value $\left\langle \left[a_{0}^{\dagger}a_{I},V\right]\right\rangle $
is negligible, we have that the mixed term described by the second
equation is also negligible --- as it can be said for the last term
in the third equation --- so that the excited part $\phi_{1}$
can be considered to evolve separately from $\phi_{0}$ in first approximation. 
Leading contributions
from $\left\langle \phi_{1}^{\dagger}\left[\phi_{1},V\right]\right\rangle $
must be those quadratic in the operators $\phi_{1}$ and $\phi_{1}^{\dagger}$,
and therefore the third equation can be approximated as
\begin{eqnarray}
\left\langle \phi_{1}^{\dagger}i\partial_{t}\phi_{1}\right\rangle \approx & \left\langle \phi_{1}^{\dagger}\left(-\frac{\nabla^{2}}{2m}\phi_{1}+2\lambda\phi_{0}^{\dagger}\phi_{0}\phi_{1}+\lambda\phi_{1}^{\dagger}\phi_{0}\phi_{0}\right)\right\rangle \, . 
\end{eqnarray}
This equation can be compared to the Bogoliubov-de Gennes equation. 
If we rewrite it in terms of the number-conserving operators, and
we consider the fact that the terms mixing the derivative of $\phi_{1}$
with $\phi_{0}$ are negligible, we can write an effective linearised
equation for $N_{0}^{-1/2}a_{0}^{\dagger}\phi_{1}$:
\begin{eqnarray}
\fl
i\partial_{t}\left(N_{0}^{-1/2}a_{0}^{\dagger}\phi_{1}\left(x\right)\right) & \approx-\frac{\nabla^{2}}{2m}\left(N_{0}^{-1/2}a_{0}^{\dagger}\phi_{1}\left(x\right)\right)+2\lambda\rho_{0}\left(x\right)\left(N_{0}^{-1/2}a_{0}^{\dagger}\phi_{1}\left(x\right)\right)
\nonumber \\
\fl
&\phantom{=}+\lambda\rho_{0}\left(x\right)e^{2i\theta_{0}\left(x\right)}\left(\phi_{1}^{\dagger}\left(x\right)a_{0}N_{0}^{-1/2}\right) \label{NumPresBdG}\, .
\end{eqnarray}
In this equation we use the functions $\rho_{0}$ and $\theta_{0}$
which are obtained from the condensed wavefunction, by writing it
as $\left\langle N_{0}\right\rangle ^{1/2}f_{0}=\rho_{0}^{1/2}e^{i\theta_{0}}$.
One can effectively assume the condensed function to be the solution
of the Gross-Pitaevskii equation, as the first corrections will be
of a lower power of $\left\langle N_{0}\right\rangle$ (and include a backreaction from this equation itself).

Assuming that $\rho_{0}$ is, at first approximation, 
homogeneous implies that the term 
$\left\langle \left[a_{0}^{\dagger}a_{I},V\right]\right\rangle $ is negligible.
If $\rho_{0}$ and $\theta_{0}$ are ultimately the same as those
obtained from the Gross-Pitaevskii equation, the same equation that
holds for the operator $\delta\phi$ can be assumed to hold for the
operator $N_{0}^{-1/2}a_{0}^{\dagger}\phi_{1}$. The solution for the mean-field description
of the condensate is therefore a general feature of the system in
studying the quantum perturbation of the condensate, not strictly
reserved to coherent states.

While having strongly related dynamical equations, the substantial
difference between the operators $\delta\phi$ of
Eq.~(\ref{Mean-Field splitting}) 
and $N_{0}^{-1/2}a_{0}^{\dagger}\phi_{1}$
is that the number-conserving operator
does not satisfy the canonical commutation
relations with its Hermitian conjugate, as we have extracted the degree
of freedom relative to the condensed state:
\begin{eqnarray}
\left[\phi_{1}\left(x\right),\phi_{1}^{\dagger}\left(y\right)\right]= & \delta\left(x,y\right)-f_{0}\left(x\right)\overline{f_{0}}\left(y\right)\, .
\end{eqnarray}
While this does not imply a significant obstruction, one must remind
that the field $\phi_{1}$ should never be treated as a canonical
quantum field. What has to be done, instead, is considering its components
with respect to the basis of time-dependent orbitals. Each mode of
the projection $\phi_{1}$ behaves as if it is a mode of a canonical
scalar quantum field in a curved spacetime. Keeping this in mind we
can safely retrieve analogue gravity.

\section{Analogue gravity with atom number conservation}\label{sec:AG-Num}
In the previous Section we discussed the equivalent of the Bogoliubov-de Gennes equation
in a number-conserving framework. Our aim in the present Section is to extend this description
to analogue gravity.

The field operators required for analogue gravity will differ from
those relative to the mean-field description, and they should be defined
considering that we have by construction removed the contribution
from the condensed $1$-particle state $f_{0}$.
The dynamical equation for the excited part in the number-conserving
formalism Eq.~\eref{NumPresBdG} appears to be the same as for the case of coherent
states (as in Eq.~\eref{Bogoliubov-de Gennes}), but instead of the field $\delta\phi$ one has $N_{0}^{-1/2}a_{0}^{\dagger}\phi_{1}$,
where we remind that $N_0=a_{0}^{\dagger} a_{0}$.

Using the Madelung representation, we may redefine the real functions
$\rho_{0}$ and $\theta_{0}$ from the condensed wavefunction $f_{0}$
and the expectation value $\left\langle N_{0}\right\rangle $. Approximating at the leading
order, we can obtain their dynamics as in the quantum Euler equations~\eref{Continuity equation}--\eref{Bernoulli equation}
\begin{eqnarray}
\left\langle N_{0}\right\rangle ^{1/2}f_{0} & =\rho_{0}e^{i\theta_{0}} \, .\label{AG-Num Mad1}
\end{eqnarray}
These functions enter in the definition of the quantum operators $\theta_{1}$
and $\rho_{1}$, which take a different expression from the usual Madelung
representation when we employ the set of number-conserving ladder operators
\begin{eqnarray}
  \theta_{1} & =-\frac{i}{2}\left\langle N_{0}\right\rangle ^{-1/2}\sum_{I\neq0}\left(\frac{f_{I}}{f_{0}}N_{0}^{-1/2}a_{0}^{\dagger}a_{I}-\frac{\overline{f_{I}}}{\overline{f_{0}}}a_{I}^{\dagger}a_{0}N_{0}^{-1/2}\right)
  \label{decomp theta1}\nonumber\\
 & =-\frac{i}{2}\left(\frac{N_{0}^{-1/2}\phi_{0}^{\dagger}\phi_{1}-\phi_{1}^{\dagger}\phi_{0}N_{0}^{-1/2}}{\left\langle N_{0}\right\rangle ^{1/2}\overline{f_{0}}f_{0}}\right)\, ,\\
\nonumber \\
\rho_{1} & =\left\langle N_{0}\right\rangle ^{1/2}\sum_{I\neq0}\left(\overline{f_{0}}f_{I}N_{0}^{-1/2}a_{0}^{\dagger}a_{I}+f_{0}\overline{f_{I}}a_{I}^{\dagger}a_{0}N_{0}^{-1/2}\right)=\label{decomp ro1}\nonumber\\
 & =\left\langle N_{0}\right\rangle ^{1/2}\left(N_{0}^{-1/2}\phi_{0}^{\dagger}\phi_{1}+\phi_{1}^{\dagger}\phi_{0}N_{0}^{-1/2}\right)\, .
\end{eqnarray}
From equations~\eref{decomp theta1} and \eref{decomp ro1} we observe
that the structure of the operators $\theta_{1}$ and $\rho_{1}$
consists of a superposition of modes, each dependent on a different
eigenfunction $f_{I}$ of the $2$-point correlation function, with a 
sum over the index $I\neq0$.

The new fields $\theta_{1}$ and $\rho_{1}$ do not satisfy the canonical commutation relations
since the condensed wavefunction $f_{0}$ is treated
separately by definition. However, these operators could be
analysed mode-by-mode, and therefore be compared in full extent to the modes
of quantum fields in curved spacetime to which they are analogous. Their modes satisfy the relations
\begin{eqnarray}
\left[\theta_{I},\rho_{J}\right] & =-i\overline{f_{I}}f_{I}\delta_{JI} \qquad \qquad \forall I,J\neq0 \label{Commutation_orbital_basis}\, .
\end{eqnarray}
Eq.~\eref{Commutation_orbital_basis} is a basis-dependent expression which can in general be found
for the fields of interest. In the simplest case of homogeneous
density of the condensate $\rho_{0}$, this commutation relations
reduce to $-i\delta_{IJ}$, and the Fourier transform provides the
tools to push the description to full extent where the indices labelling
the functions are the momenta $k$.

The equations for analogue gravity are found under the usual assumptions
regarding the quantum pressure, \emph{i.e.} the space gradients of the atom
densities are assumed to be small. When considering homogeneous condensates
this requirement is of course satisfied. In non-homogeneous
condensates we require
\begin{eqnarray}
\nabla\left(\rho_{0}^{-1}\left(\nabla\rho_{0}\right)\right) & \ll4m\lambda\rho_{0}\, ,\label{assumption1}\\
\nabla\left(\rho_{0}^{-1}\left(\nabla\rho_{1}\right)\right) &
\ll4m\lambda\rho_{1}\, .\label{assumption2}
\end{eqnarray}
Making the first assumption~\eref{assumption1}, the effective
coupling constant $\widetilde{\lambda}$ is a global feature of the
system with no space dependence. This means that all the
inhomogeneities of the system are encoded in the velocity of the superfluid,
the gradient of the phase of the condensate. As stated before, the
continuity equation can induce inhomogeneities in the density
if there are initial inhomogeneities in the phase, but for sufficiently
short intervals of time the assumption is satisfied. Another effect
of the first assumption~\eref{assumption1} is that the term
$\int \mathrm{d}x \overline{f_{I}}\overline{f_{0}}f_{0}f_{0}$
is negligible. The more $\rho_{0}$ is homogeneous, the closest
this integral is to vanishing, making the description more consistent.
The second assumption~\eref{assumption2} is a general requirement in
analogue gravity, needed to have local Lorentz symmetry, and therefore
a proper Klein-Gordon equation for the field $\theta_{1}$. When $\rho_{0}$
is homogeneous this approximation means considering only small momenta,
for which we have the usual dispersion relation.

Under these assumptions, the usual equations for analogue gravity are
obtained
\begin{eqnarray}
\fl
\quad \rho_{1} & =-\frac{1}{\lambda}\left(\left(\partial_{t}\theta_{1}\right)+\frac{1}{m}\left(\nabla\theta_{0}\right)\left(\nabla\theta_{1}\right)\right)\, ,\\
\fl
\left(\partial_{t}\rho_{1}\right) & =-\frac{1}{m}\nabla\left(\rho_{1}\left(\nabla\theta_{0}\right)+\rho_{0}\left(\nabla\theta_{1}\right)\right)\\
\fl
 & \Downarrow \nonumber \\
\fl
\qquad 0 & =\partial_{t}\left(-\frac{1}{\lambda}\left(\partial_{t}\theta_{1}\right)-\frac{\delta^{ij}}{m\lambda}\left(\nabla_{j}\theta_{0}\right)\left(\nabla_{i}\theta_{1}\right)\right)\nonumber \\ \*
\fl
 & \phantom{=}+\nabla_{j}\left(-\frac{\delta^{ij}}{m\lambda}\left(\nabla_{i}\theta_{0}\right)\left(\partial_{t}\theta_{1}\right)+\left(\frac{\delta^{ij}\rho_{0}}{m}-\frac{1}{\lambda}\frac{\delta^{il}}{m}\frac{\delta^{jm}}{m}\left(\nabla_{l}\theta_{0}\right)\left(\nabla_{m}\theta_{0}\right)\right)\left(\nabla_{i}\theta_{1}\right)\right)
\end{eqnarray}
so that $\theta_{1}$ is the analogue of a scalar massless field in curved
spacetime. However, the operator $\theta_{1}$ is intrinsically unable
to provide an exact full description of a massless field since it
is missing the mode $f_{0}$. Therefore, the operator $\theta_{1}$ 
is best handled when considering the propagation of its constituent modes, 
and relating them to those of the massless field. 

The viability of this description as a good analogue gravity setup
is ensured, ultimately, by the fact that the modes of $\theta_{1}$,
\emph{i.e.} the operators describing the excited part of the atomic field,
have a closed dynamics. The most important feature in the effective
dynamics of the number-conserving operators $N_{0}^{-1/2}a_{0}^{\dagger}a_{I}$,
as described in equation Eq.~\eref{NumPresBdG}, is that its time derivative can
be written as a composition of the same set of number operators, and
this enables the analogue model.

In the following we are going to stick to the case of an homogenous condensate,
which is arguably the most studied case in analogue 
gravity.
The description is enormously simplified 
by the fact that the gradient of the condensed wavefunction vanishes, since 
$f_{0}=V^{-1/2}$, meaning that the condensed state is fully described by the 
state of null momentum $k=0$.
In a homogeneous BEC all the time-dependent 
orbitals are labeled by the momenta they carry, and at every moment
in time we can apply the same Fourier transform to transform the differential equations in the 
space of coordinates to algebraic equations in the space of momenta. We expect
that the number-conserving treatment of the inhomogenous condensate follows
along the same lines, albeit being technically more complicated.

\section{Simulating cosmology in number-conserving analogue gravity }\label{sec:SC}

With an homogeneous condensate we can simulate a cosmology with a 
scale factor changing in time --- as long as we can control and modify in time the 
strength of the $2$-body interaction $\lambda$ --- and we can verify the prediction 
of quantum field theory in curved spacetime that in an expanding universe one 
should observe a cosmological particle creation. In this setup there is no 
ambiguity in approximating the mixed term of the interaction potential, as 
discussed in the Subsection~\eref{connection}.

To further proceed, we apply the usual transformation to pass from the Bogoliubov description of the 
atomic system to the setup of analogue gravity, and we then proceed considering 
number-conserving operators.
It is convenient to adopt a compact notation for the condensate wavefunction and 
its approximated dynamics, as discussed previously in equations~\eref{Continuity equation}--\eref{Bernoulli equation} and in Eq.~\eref{AG-Num Mad1} 
\begin{eqnarray}
f_{0}\left(x\right)\left\langle N_{0}\right\rangle ^{1/2}&\equiv \phi_{0}=\rho_{0}e^{i\theta_{0}}\, ,\\
\qquad \qquad \partial_{t}\rho_{0}&= 0\, ,\\
\qquad \qquad \partial_{t}\theta_{0}&= -\lambda\rho_{0}\, .
\end{eqnarray}
To study the excitations described by the operator $\theta_{1}\left(x\right)$ we need the basis 
of time-dependent orbitals, which in the case of a homogeneous condensate is given by the plane waves,
the set of orthonormal functions which define the Fourier transform and are labeled by the momenta.
By Fourier transforming the operator $\phi_{1}\left(x\right)$, orthogonal 
to the condensate wavefunction, 
we have
\begin{eqnarray}
  \qquad \quad \delta\phi_{k} & \equiv \int\frac{\mathrm{d}x}{\sqrt{V}}e^{-ikx}N_{0}^{-1/2}a_{0}^{\dagger}\phi_{1}\left(x\right) \nonumber
  \\
  &= \int\frac{\mathrm{d}x}{\sqrt{V}}e^{-ikx}N_{0}^{-1/2}a_{0}^{\dagger}\sum_{q\neq 0}\frac{e^{iqx}}{\sqrt{V}}a_{q} \nonumber
  \\
&= N_{0}^{-1/2}a_{0}^{\dagger}a_{k}\, ,\label{def_nc_deltaphi}\\
\left[\delta\phi_{k},\delta\phi_{k^{\prime}}^{\dagger}\right]&= \delta_{k,k^{\prime}} 
\qquad \qquad \forall k,k^{\prime}\neq0 \label{comm_delta_delta} \, .
\end{eqnarray}
Notice that with the notation of ~\eref{alphaa}, $\delta\phi_{k}$ would be
just $\alpha_{k}$ and one sees the dependence on the condensate operator
$a_0$. 

Following the same approach discussed in the Sec.~\eref{sec:AG-Num}, we define $\theta_{k}$ and 
$\rho_{k}$. These number-conserving operators are labeled with a non-vanishing 
momentum and act in the atomic Fock space, in a superposition of two operations, 
extracting momentum $k$ from the state or introducing momentum $-k$ to it. 
All the following relations are defined for $k,k^{\prime}\neq0$
\begin{eqnarray}
  \qquad \qquad \qquad \theta_{k}&= -\frac{i}{2}\left(\frac{\delta\phi_{k}}{\phi_{0}}-\frac{\delta\phi_{-k}^{\dagger}}{\overline{\phi_{0}}}\right) \label{theta_from_delta}\, ,\\
 \qquad \qquad \qquad \rho_{k}&= \rho_{0}\left(\frac{\delta\phi_{k}}{\phi_{0}}+\frac{\delta\phi_{-k}^{\dagger}}{\overline{\phi_{0}}}\right)\, ,\\
\qquad \qquad \left[\theta_{k},\rho_{k^{\prime}}\right]&= -i\left[\delta\phi_{k},\delta\phi_{-k^{\prime}}^{\dagger}\right]=-i\delta_{k,-k^{\prime}}\, , \label{Commutation_theta_rho_num}\\
 \left\langle \delta\phi_{k}^{\dagger}\left(t\right)\delta\phi_{k^{\prime}}\left(t\right)\right\rangle &= \delta_{k,k^{\prime}}\left\langle N_{k}\right\rangle\, .\label{comm_delta_phi}
\end{eqnarray}
Again we remark that these definitions of $\theta_{k}$ and $\rho_{k}$ do 
not provide, through an inverse Fourier transform of these operators, a couple of 
conjugate real fields $\theta_{1}\left(x\right)$  and $\rho_{1} \left( x \right)$ with 
the usual commutation relations as in Eq.\eref{canonical_comm_rel}, because they are not relative to 
a set of functions that form a complete basis of the 1-particle Hilbert space, as the 
mode $k=0$ is not included.
But these operators, describing each mode with $k\neq0$, can 
be studied separately and they show the same behaviour of the 
components of a quantum field in curved spacetime: the commutation relations in 
Eq.~\eref{Commutation_theta_rho_num} are the same as those that are satisfied by the components of a quantum scalar field.

From the Bogoliubov-de Gennes equation~\eref{NumPresBdG} we get the two 
coupled dynamical equations for $\theta_{k}$ and $\rho_{k}$
\begin{eqnarray}
\partial_{t}\theta_{k}&= -\frac{1}{2}\left(\frac{k^{2}}{2m}+2\lambda\rho_{0}\right)\frac{\rho_{k}}{\rho_{0}}\, ,\label{Dynamics th_k}\\
\partial_{t}\frac{\rho_{k}}{\rho_{0}}&= \frac{k^{2}}{m}\theta_{k}\, .\label{Dynamics ro_k}
\end{eqnarray}
Combining these gives the analogue Klein-Gordon equation 
for each mode $k\neq0$ 
\begin{eqnarray}
\partial_{t}\left(-\frac{1}{\lambda\rho_{0}+\frac{k^{2}}{4m}}\left(\partial_{t}\theta_{k}\right)\right)&= \frac{k^{2}}{m}\theta_{k}\, .\label{Klein-Gordon}
\end{eqnarray}
In this equation the term due to quantum pressure is retained
for convenience, since the 
homogeneity of the condensed state makes it easy to maintain it in
the description. It modifies the dispersion 
relation and breaks Lorentz symmetry,
but the usual expression is found in the limit
$\frac{k^{2}}{2m}\ll2\lambda\rho_{0}$.

When the quantum pressure is neglected, the analogue metric tensor is
\begin{eqnarray}
g_{\mu\nu}\mathrm{d}x^{\mu}\mathrm{d}x^{\nu}&=\sqrt{\frac{\rho_{0}}{m\lambda}}\left(-\frac{\lambda\rho_{0}}{m}\mathrm{d}t^{2}+\delta_{ij}\mathrm{d}x^{i}\mathrm{d}x^{j}\right) \, .\label{cosmological_metric_tensor}
\end{eqnarray}
This metric tensor is clearly analogous to that of a cosmological spacetime, where
the evolution is given by the time dependence of the coupling constant $\lambda$.
This low-momenta limit is the regime in which we are mostly interested, because 
when these conditions are realised the quasi-particles,
the excitations of the field 
$\theta_{k}$, behave most similarly to particles in a curved spacetime 
with local Lorentz symmetry.

\subsection{Cosmological particle production}
We now consider a setup for which the coupling constant
varies from an initial value 
$\lambda$ to a final value $\lambda^{\prime}$ through a transient phase.
$\lambda$ is assumed asymptotically constant
for both $t\rightarrow\pm\infty$. This setup has been studied in the Bogoliubov
approximation in~\cite{Barcelo:2003wu,PhysRevA.76.033616,Weinfurtner_2009,Carusotto2010,PhysRevLett.109.220401} 
and can be experimentally realised with \emph{e.g.} via Feshbach resonance.
For one-dimensional Bose gases, where significant corrections to the Bogoliubov approximation are expected
far from the weakly interacting limit, a study of the large time evolution of correlations was presented
in~\cite{Sotiriadis_2014}. Here our aim is to study the effect of the variation of the coupling constant
in the number-conserving framework. 

There will 
be particle creation and the field in general takes the expression
\begin{eqnarray}
\theta_{k}\left(t\right)&= \frac{1}{\mathcal{N}_{k}\left(t\right)}\left(e^{-i\Omega_{k}\left(t\right)}c_{k}+e^{i\Omega_{-k}\left(t\right)}c_{-k}^{\dagger}\right)\, , \label{c_from_theta}
\end{eqnarray}
where the operators $c_{k}$ are the creation and destruction operators for the 
quasi-particles at $t\rightarrow-\infty$. For the time $t\rightarrow+\infty$
there will 
be a new set of operators $c^{\prime}$
\begin{eqnarray}
\theta_{k}\left(t\rightarrow-\infty\right)&=\frac{1}{\mathcal{N}_{k}}\left(e^{-i\omega_{k}t}c_{k}+e^{i\omega_{k}t}c_{-k}^{\dagger}\right) \, ,\\
\theta_{k}\left(t\rightarrow+\infty\right)&=\frac{1}{\mathcal{N}_{k}^{\prime}}\left(e^{-i\omega_{k}^{\prime}t}c_{k}^{\prime}+e^{i\omega_{k}^{\prime}t}c_{-k}^{\prime\dagger}\right) \, .\label{theta_k final}
\end{eqnarray}
From these equations, in accordance with Eq.~\eref{Dynamics th_k}, 
we obtain
\begin{equation}
  \rho_{k}=-\frac{2\rho_{0}}{\frac{k^{2}}{2m}+2\lambda\rho_{0}}\partial_{t}\theta_{k}
\end{equation}
and the two following asymptotic expressions for $\rho_{k}$:
\begin{eqnarray}
\rho_{k}\left(t\rightarrow-\infty\right)&=\frac{2i\omega_{k}\rho_{0}}{\frac{k^{2}}{2m}+2\lambda\rho_{0}}\frac{1}{\mathcal{N}_{k}}\left(e^{-i\omega_{k}t}c_{k}-e^{i\omega_{k}t}c_{-k}^{\dagger}\right)\, ,\\
\rho_{k}\left(t\rightarrow+\infty\right)&=\frac{2i\omega_{k}^{\prime}\rho_{0}}{\frac{k^{2}}{2m}+2\lambda^{\prime}\rho_{0}}\frac{1}{N_{k}^{\prime}}\left(e^{-i\omega_{k}^{\prime}t}c_{k}^{\prime}-e^{i\omega_{k}^{\prime}t}c_{-k}^{\prime\dagger}\right)\, .
\end{eqnarray}
With the previous
expressions for $\theta_{k}$ and $\rho_{k}$ and imposing the commutation
relations in
Eq.~\eref{Commutation_theta_rho_num}, we
retrieve the energy spectrum
$\omega_{k}= \sqrt{\frac{k^{2}}{2m}\left(\frac{k^{2}}{2m}+2\lambda\rho_{0}\right)}$ as expected and the (time-dependent) 
normalization prefactor $\mathcal{N}$:
\begin{equation}
\mathcal{N}_{k}= \sqrt{4\rho_{0}\sqrt{\frac{\frac{k^{2}}{2m}}{\frac{k^{2}}{2m}+2\lambda\rho_{0}}}}\, .
\end{equation}
The expected
commutation relations for the operators $c$ and $c^{\prime}$ are found
(again not including the mode $k=0$):
\begin{eqnarray}
\quad 0&=\Bigl[c_{k},c_{k^{\prime}}\Bigr]=\Bigl[c_{k}^{\prime},c_{k^{\prime}}^{\prime}\Bigr]\, ,
\\\delta_{k,k^{\prime}}&=\Bigl[c_{k},c_{k^{\prime}}^{\dagger}\Bigr]=\Bigl[c_{k}^{\prime},c_{k^{\prime}}^{\prime\dagger}\Bigr] \label{comm_c_c}\, .
\end{eqnarray}
It is found
\begin{equation}
c_{k}^{\prime}= \cosh\Theta_{k}c_{k}+\sinh\Theta_{k}e^{i\varphi_{k}}c_{-k}^{\dagger}\label{bogoliubov_quasi-particles}
\end{equation}
with 
\begin{eqnarray} 
\quad \cosh\Theta_{k}&= \cosh\Theta_{-k} \, , \label{Condition Theta} \\
\sinh\Theta_{k}e^{i\varphi_{k}}&= \sinh\Theta_{-k}e^{i\varphi_{-k}} \label{Condition varphi} \, .
\end{eqnarray}

The initial state in which we are interested is the vacuum of quasi-particles,
so that each quasi-particle destruction operators $c_{k}$ annihilates the initial state~\footnote{To make contact with the standard Bogoliubov
  approximation, if there one denotes
  by $\gamma_k$ the quasi-particles one has that the $\gamma_k$ are a
  combination of the atom operators $a_{k}$, $a_{-k}$ of the form
  $\gamma_k=u_k a_k+v_k a_{-k}^{\dagger}$~\cite{Pethick20081}. Correspondingly,
  in the number-conserving formalism the quasi-particle operators $c_k$ are a
combination of the atom operators $\delta \phi_k \equiv \alpha_{k}$,
$\delta\phi_{-k} \equiv \alpha_{-k}$.}: 
\begin{eqnarray}
c_{k}\left| in \right\rangle\equiv&0  \qquad \qquad \forall k\neq0\, .\label{vacuum_condition}
\end{eqnarray}
To realise this initial condition we should impose constraints,
in principle, on every correlation function.
We focus on the 2-point correlation functions $\left \langle \delta\phi^{\dagger}\delta\phi\right\rangle$ and $\left\langle \delta\phi\delta\phi\right\rangle$. 
In particular, the first of the two determines the number of atoms with momentum $k$
in the initial state:
\begin{eqnarray}
\left\langle \delta\phi_{k}^{\dagger}\delta\phi_{k}\right\rangle &= \left\langle a_{k}^{\dagger}a_{0}N_{0}^{-1}a_{0}^{\dagger}a_{k}\right\rangle =\left\langle a_{k}^{\dagger}a_{k}\right\rangle =\left\langle N_{k}\right\rangle\, . 
\end{eqnarray}
In order for the state to be condensed with respect to the
state with momentum $0$, it 
must be that $\left\langle N_{k}\right\rangle \ll\left\langle N_{0}\right\rangle =\rho_{0}V$.
When the vacuum condition Eq.~\eref{vacuum_condition} holds, the 2-point correlation
functions can be easily evaluated to be:
\begin{eqnarray}
\left\langle \delta\phi_{k}^{\dagger}\delta\phi_{k^{\prime}}\right\rangle &
=\left(\frac{1}{2}\frac{\frac{k^{2}}{2m}+\lambda\rho_{0}}{\sqrt{\frac{k^{2}}{2m}\left(\frac{k^{2}}{2m}+2\lambda\rho_{0}\right)}}-\frac{1}{2}\right)\delta_{k,k^{\prime}}\nonumber\\
&\approx\frac{1}{4}\sqrt{\frac{2\lambda\rho_{0}}{\frac{k^{2}}{2m}}}\delta_{k,k^{\prime}}\, ,\\
\left\langle \delta\phi_{-k}\delta\phi_{k^{\prime}}\right\rangle &=-\frac{e^{2i\theta_{0}}}{4}\frac{2\lambda\rho_{0}}{\sqrt{\frac{k^{2}}{2m}\left(\frac{k^{2}}{2m}+2\lambda\rho_{0}\right)}}\delta_{k,k^{\prime}}\nonumber\\
&\approx-e^{2i\theta_{0}}\left\langle \delta\phi_{k}^{\dagger}\delta\phi_{k^{\prime}}\right\rangle \, ,
\end{eqnarray}
where in the last line we have used $\frac{k^{2}}{2m}\ll2\lambda\rho_{0}$, the limit in which the quasi-particles 
propagate in accordance with the analogue metric Eq.~\eref{cosmological_metric_tensor}, and one has to keep into account that the phase of the condensate is time dependent and consequently the last correlator is oscillating.

We now see that the conditions of condensation $\left\langle N_{k}\right\rangle \ll\left\langle N_{0}\right\rangle$ and of low-momenta translate into
\begin{equation}
\frac{2\lambda\rho_{0}}{16\left\langle N_{0}\right\rangle ^{2}} \ll\frac{k^{2}}{2m}\ll2\lambda\rho_{0}\, .
\end{equation}
The range of momenta that should be considered is
therefore set by the number of condensate atoms, the physical dimension
of the atomic system and the strength of the $2$-body interaction.

The operators $\theta_{k}$ satisfying Eq.~\eref{Klein-Gordon} --- describing the excitations of quasi-particles over
a BEC --- are analogous to the components of a scalar quantum field in a cosmological spacetime.
In particular, if we consider a cosmological metric given in the usual form
\begin{eqnarray}
g_{\mu\nu}\mathrm{d}x^{\mu}\mathrm{d}x^{\nu}&=-\mathrm{d}\tau^{2}+a^{2}\delta_{ij}\mathrm{d}x^{i}\mathrm{d}x^{j}\, ,
\end{eqnarray}
the analogy is realised for a specific relation
between the coupling $\lambda \left( t \right)$ 
and the scale factor $a \left( \tau \right)$,
which then induces the relation between the laboratory time $t$ and 
the cosmological time $\tau$. These relations are given by
\begin{eqnarray}
a\left(\tau\left(t\right)\right)&=\left(\frac{\rho}{m\lambda\left(t\right)}\right)^{1/4}\frac{1}{C}\, ,\\
\qquad \mathrm{d}\tau&=\frac{\rho}{ma\left(\tau\left(t\right)\right)}\frac{1}{C^{2}}\mathrm{d}t\,,
\end{eqnarray}
for an arbitrary constant $C$.

In cosmology the evolution of the scale factor
leads to the production of particles by cosmological particle creation, 
as implied by the Bogoliubov transformation relating the operators
which, at early and late times, 
create and destroy the quanta we recognise as particles.
The same happens for the quasi-particles over the
condensate, as discussed in Sec.~\eref{sec:SC},
because the coupling $\lambda$ is time-dependent and
the definition itself of quasi-particles
changes from initial to final time. 
The ladder operators associated to these quasi-particles are related to each other by the Bogoliubov
transformation introduced in Eq.~\eref{bogoliubov_quasi-particles}, fully defined by
the parameters $\Theta_{k}$ and $\varphi_{k}$ (which must also satisfy Eq.~\eref{Condition Theta} and Eq.~\eref{Condition varphi}).

\subsection{Scattering operator}

The exact expressions of $\Theta_{k}$ and $\varphi_{k}$ 
depend on the behaviour of $\lambda\left(t\right)$,
which is a function of the cosmological scale parameter and is therefore different for each cosmological model. 
They can in general be evaluated with the well
established methods used in quantum field theory
in curved spacetimes~\cite{Birrell:1982ix}.
In general it is found that $\cosh\Theta_{k}>1$, as the value
$\cosh\Theta_{k}=1$ (\emph{i.e.} $\sinh\Theta_{k}=0$) is restricted to the case in which $\lambda$
is a constant for the whole evolution, and the analogue spacetime
is simply flat.

The unitary operator describing the evolution from initial to final time
is $U\left(t_{out},t_{in}\right)$; when $t_{out}\rightarrow+\infty$
and $t_{in}\rightarrow-\infty$, the operator $U$,
is the scattering operator $S$.
This is exactly the operator acting on the quasi-particles,
defining the Bogoliubov transformation in which we are interested
\begin{eqnarray}
c_{k}^{\prime}&= S^{\dagger}c_{k}S \, .
\end{eqnarray}
The behavior of $c_{k}^{\prime}$, describing the quasi-particles at late times, 
can therefore be understood from the behaviour
of the initial quasi-particle operators $c_{k}$ 
when the expression of the scattering operator is known. 
In particular the phenomenon of cosmological particle creation
is quantified considering
the difference between the expectation value of the number operator
of quasi-particles with given momentum at early times
and the expectation values of the number operator of quasi-particles
of same momentum at late times.

Consider as initial state the vacuum of quasi-particles at early times,
satisfying the condition Eq.~\eref{vacuum_condition}. It is analogue
to a Minkowski vacuum, and
the evolution of the coupling $\lambda \left( t\right)$ induces
a change in the definition of quasi-particles. We find that,
of course, the state is
not a vacuum with respect to the final quasi-particles $c^{\prime}$. It is
\begin{equation}
\fl
\quad S^{\dagger}c_{k}^{\dagger}c_{k}S= c_{k}^{\prime\dagger}c_{k}^{\prime}=\left(\cosh\Theta_{k}c_{k}^{\dagger}+\sinh\Theta_{k}e^{-i\varphi_{k}}c_{-k}\right)\left(\cosh\Theta_{k}c_{k}+\sinh\Theta_{k}e^{i\varphi_{k}}c_{-k}^{\dagger}\right)
\end{equation}
and
\begin{equation}
\left\langle S^{\dagger}c_{k}^{\dagger}c_{k}S\right\rangle = 
\sinh^{2}\Theta_{k}\left\langle c_{-k}c_{-k}^{\dagger}\right\rangle =\sinh^{2}\Theta_{k}>0 \, .
\end{equation}

We are interested in the impact that the evolution of the quasi-particles
have on the atoms. The system
is fully characterised by the initial conditions and the
Bogoliubov transformation: we have
the initial occupation numbers, the range
of momenta which we should consider and
the relation between initial and final quasi-particles.

What is most significant is that the quasi-particle dynamics affects
the occupation number of the atoms.
Considering that for $t$ sufficiently large we are already in the final regime,
the field takes the following values:
\begin{eqnarray}
\fl
\delta\phi_{k}\left(t\rightarrow-\infty\right)&= i\rho_{0}^{1/2}e^{i\theta_{0}\left(t\right)}\frac{1}{\mathcal{N}_{k}}\left(\left({\cal F}_k+1\right)e^{-i\omega_{k}t}c_{k}-\left({\cal F}_k-1\right)e^{i\omega_{k}t}c_{-k}^{\dagger}\right)\\
\fl
\delta\phi_{k}\left(t\rightarrow+\infty\right)&= i\rho_{0}^{1/2}e^{i\theta_{0}\left(t\right)}\frac{1}{\mathcal{N}_{k}^{\prime}}\left(\left({\cal F}_{k}^{\prime}+1\right)e^{-i\omega_{k}^{\prime}t}c_{k}^{\prime}-\left({\cal F}_{k}^{\prime}-1\right)e^{i\omega_{k}^{\prime}t}c_{-k}^{\prime\dagger}\right)\\
\end{eqnarray}
where ${\cal F}_k \equiv \frac{\omega_{k}}{\frac{k^{2}}{2m}+2\lambda\rho_{0}}$ and
${\cal F}_{k}^{\prime} \equiv \frac{\omega_{k}^{\prime}}{\frac{k^{2}}{2m}+2\lambda^{\prime}\rho_{0}}$, with
$\omega_{k}^{\prime}=\sqrt{\frac{k^{2}}{2m}\left(\frac{k^{2}}{2m}+2\lambda^{\prime}\rho_{0}\right)}$. One finds
\begin{eqnarray}
\fl \left\langle \delta\phi_{k}^{\dagger}\left(t\right)\delta\phi_{k}\left(t\right)\right\rangle &= 
\frac{\frac{k^{2}}{2m}+\lambda^{\prime}\rho_{0}}{2\omega_{k}^{\prime}}\cosh\left(2\Theta_{k}\right)-\frac{1}{2}
+\frac{\lambda^{\prime}\rho_{0}\sinh\left(2\Theta_{k}\right)}{2\omega_k^{\prime}}\cos\left(2\omega_k^{\prime}t-\varphi_{k}\right)\label{Extraction}\, .
\end{eqnarray}
In Eq.\eref{Extraction} the last term is oscillating symmetrically around $0$ ---
meaning that the atoms will leave and rejoin the condensate periodically in time --- while the first two are stationary. 

An increase in the value of the coupling $\lambda$ has therefore
deep consequences. It appears explicitly in the prefactor
and more importantly it affects the hyperbolic functions
$\cosh\Theta_{k}>1$, which implies that the mean value is
larger than the initial one,
differing
from the equilibrium value corresponding to the vacuum of quasi-particles.

This result is significant because it explicitly shows that the quasi-particle
dynamics influences the underlying structure of atomic particles.
Even assuming that the backreaction of the quasi-particles on
the condensate is negligible for the dynamics of the quasi-particles
themselves, the mechanism of extraction of atoms from the condensate fraction is effective and
increases the depletion (as also found in the standard Bogoliubov approach). This extraction mechanism can be evaluated
in terms of operators describing the quasi-particles,
that can be defined \emph{a posteriori}, without notion of the operators
describing the atoms.

The fact that analogue gravity can be reproduced in
condensates independently from the use of
coherent states enhances the validity of the discussion.
It is not strictly necessary that we have a coherent state
to simulate the effects of curvature with quasi-particles,
but in the more general case of condensation,
the condensed wavefunction provides a support for the propagation of
quasi-particles. From an analogue gravity point of view,
its intrinsic role is that of the seed for
the emergence of the analogue scalar field.

\section{Squeezing and quantum state structure}\label{sec:SQSS}

The Bogoliubov transformation in Eq.~\eref{bogoliubov_quasi-particles} leading to the quasi-particle production describes
the action of the scattering operator on the ladder operators, relating the operators at early and late times. 
The linearity of this transformation is obtained by the linearity of the dynamical equation for the quasi-particles, 
which is particularly simple in the case of homogeneous condensate.

The scattering operator $S$ is unitary by definition, as it is easily checked by its action on the operators $c_{k}$.
Its full expression can be found from the Bogoliubov transformation, finding the generators of the transformation
when the arguments of the hyperbolic functions, the parameters $\Theta_{k}$, are infinitesimal:
\begin{equation}
  S^{\dagger}c_{k}S=c_{k}^{\prime}= \cosh\Theta_{k}c_{k}+\sinh\Theta_{k}e^{i\varphi_{k}}c_{-k}^{\dagger}\, .
\end{equation}
It follows
\begin{equation}
S = \exp\left(\frac{1}{2}\sum_{k\neq0}\left(-e^{-i\varphi_{k}}c_{k}c_{-k}
+e^{i\varphi_{k}}c_{k}^{\dagger}c_{-k}^{\dagger}\right)\Theta_{k}\right)\, . \label{Explicit scattering operator}
\end{equation}
The scattering operator is particularly simple, and takes the peculiar expression that is required for producing squeezed states.
This is the general functional expression which is found in cosmological particle creation and in its analogue gravity counterparts,
whether they are realised in the usual Bogoliubov framework or in its number-conserving reformulation. As discussed previously,
the number-conserving formalism is more general, reproduces the usual case when the state is an eigenstate of the destruction operator $a_{0}$, and includes the notion that the excitations of the condensate move condensate atoms to the excited part.

The expression in Eq.~\eref{Explicit scattering operator} has been found
under the hypothesis that the mean value of the operator $N_{0}$ is macroscopically 
larger than the other occupation numbers. Instead of using the quasi-particle ladder operators, $S$ can be rewritten easily in terms 
of the atom operators. In particular, we remind that the time-independent operators $c_{k}$ depend on the condensate operator $a_0$
and can be defined
as compositions of number-conserving atom operators $\delta\phi_{k}\left(t\right)$ and $\delta\phi_{-k}^{\dagger}\left(t\right)$ 
defined in Eq.~\eref{def_nc_deltaphi}. At any time there will be a transformation from a set of operators
to the other. It is significant that the operators $c_{k}$ commute with the operators $N_{0}^{-1/2}a_{0}^{\dagger}$ 
and $a_{0}N_{0}^{-1/2}$, which are therefore conserved in time
(as long as the linearised dynamics for $\delta\phi_{k}$ is a good approximation)
\begin{eqnarray}
\left[\phi_{1}\left( x \right),N_{0}^{-1/2}\left(t\right)a_{0}^{\dagger}\left(t\right)\right]&= 0\, ,\\
&\Downarrow\nonumber \\
\qquad \left[c_{k},N_{0}^{-1/2}\left(t\right)a_{0}^{\dagger}\left(t\right)\right]&= 0 \, ,\\
&\Downarrow\nonumber \\
\qquad \left[S,N_{0}^{-1/2}\left(t\right)a_{0}^{\dagger}\left(t\right)\right]&= 0 \, .
\end{eqnarray}
The operator $S$ cannot have other terms apart for those in Eq.\eref{Explicit scattering operator},
even if it is defined for its action on the 
operators $c_{k}$, and therefore on a set of functions which is not a complete basis of the 1-particle Hilbert space.
Nevertheless the notion of number conservation implies its action on the condensate and on the operator $a_{0}$.

One could investigate whether it is possible
to consider a more general expression with additional terms depending only on $a_{0}$ and $a_{0}^\dagger$, \emph{i.e.} assuming 
the scattering operator to be
\begin{eqnarray}
S&= \exp\left(\frac{1}{2}\sum_{k\neq 0}\left(Z_{k}c_{k}c_{-k}+c_{k}^{\dagger}c_{-k}^{\dagger}Z_{-k}^{\dagger}\right)+G_{0}\right)\, ,
\end{eqnarray}
where we could assume that the coefficients of the quasi-particle operators are themselves
depending on only $a_{0}$ and $a_{0}^{\dagger}$,
and so $G_{0}$. But the requirement that $S$ commutes with the total number of atoms $N$ implies that so do its generators,
and therefore $Z$ and $G_{0}$ must be functionally dependent on $N_{0}$, and not on $a_{0}$ and $a_{0}^{\dagger}$ separately,
since they do not conserve the total
number. Therefore it must hold
\begin{eqnarray}
0&= \left[\left(\frac{1}{2}\sum_{k\neq 0}\left(Z_{k}c_{k}c_{-k}+c_{k}^{\dagger}c_{-k}^{\dagger}Z_{-k}^{\dagger}\right)+G_{0}\right),N\right]\, .
\end{eqnarray}
The only expressions in agreement with the linearised dynamical equation for $\delta\phi$ imply that $Z$ and $G_{0}$ are multiple of the
identity, otherwise they would modify the evolution of the operators $\delta\phi_{k}=N_{0}^{-1/2}a_{0}^{\dagger}a_{I}$, as they do not
commute with $N_{0}$.
This means that that corrections to the scattering operator are possible only involving higher-order corrections (in terms of $\delta\phi$).

The fact that the operator $S$ as in Eq.~\eref{Explicit scattering operator} is the only number-conserving operator 
satisfying the dynamics is remarkable because it emphasises that the production of quasi-particles is a phenomenon 
which holds only in terms of excitations of atoms from the condensate to the excited part, with the number of 
transferred atoms evaluated in the previous subsection. The expression of the scattering operator shows that the 
analogue gravity system produces states in which the final state presents squeezed quasi-particle states, but the occurrence 
of this feature in the emergent dynamics happens only introducing correlations in the condensate, with each quanta of the analogue field $\theta_{1}$ entangling atoms in the condensate with atoms in the excited part.

The quasi-particle scattering operator obtained in the number-conserving framework is functionally equivalent to 
that in the usual Bogoliubov description, and the difference between the two appears when considering the atom 
operators, depending on whether $a_{0}$ is treated for its quantum nature or it is replaced with the number 
$\bigl\langle N_{0}^{1/2}\bigr\rangle$.
This reflects that the dynamical equations are functionally the same when the expectation value 
$\bigl\langle N_{0}\bigr\rangle$ is macroscopically large.

There are no requirements on the initial density matrix of the state, and
it is not relevant whether the state is a coherent superposition of infinite states with different
number of atoms or it is a pure state with a fixed number of atoms in the same 1-particle state.
The quasi-particle description holds the same and it provides the same predictions. This is useful
for implementing analogue gravity systems, but also
a strong hint in interpreting the problem of information loss.
When producing quasi-particles in analogue gravity one can, in first approximation,
reconstruct the initial expectation values of the excited states, and push the description to include the backreaction on the condensate.
What we are intrinsically unable to do is reconstruct the entirety of the initial atom quantum state, \emph{i.e.}
how the condensate is composed. 

We know that in analogue gravity the evolution is unitary, the final state is uniquely determined by the initial state.
Knowing all the properties of the final state
we could reconstruct the initial state, and yet the intrinsic inability to infer all the properties of the condensate atoms from
the excited part shows that
the one needs to access the full correlation properties of the condensate atoms with the quasi-particles
to fully appreciate (and retrieve) the unitarity of the evolution.

\subsection{Correlations}
In the previous Section we made the standard choice of considering as initial state the quasi-particle vacuum.
To characterise it with respect to the atomic degrees of freedom, the quasi-particle ladders operators have to
be expressed as compositions of the number-conserving atomic operators, manipulating Eq.~\eref{theta_from_delta} and Eq.~\eref{c_from_theta}. 

By definition, at any time, both sets of operators satisfy the canonical commutation relations~\eref{comm_delta_phi} and~\eref{comm_c_c}
$\forall k,k^{\prime}\neq0$. 
Therefore, it must exist a Bogoliubov transformation linking the quasi-particle and the number-conserving operators
which will in general be written as
\begin{eqnarray}
c_{k}=	e^{-i\alpha_{k}}\cosh\Lambda_{k}\delta\phi_{k}+e^{i\beta_{k}}\sinh\Lambda_{k}\delta\phi_{-k}^{\dagger}\label{Bogoliubov_delta_c}\, .\end{eqnarray}
The transformation is defined through a set of functions $\Lambda_{k}$, constant in the stationary case,
and the phases $\alpha_{k}$ and $\beta_{k}$, inheriting their time dependence from the atomic operators.
These functions can be obtained from Eq.~\eref{theta_from_delta} and Eq.~\eref{c_from_theta}:
\begin{eqnarray}
\cosh\Lambda_{k}	=\left(\frac{\omega_{k}+\left(\frac{k^{2}}{2m}+2\lambda\rho_{0}\right)}{4\omega_{k}}\right)\frac{\mathcal{N}_{k}}{\phi_{0}}\, .\end{eqnarray}
If the coupling changes in time, the quasi-particle operators during the transient are defined knowing the solutions of the Klein-Gordon equation.
With the Bogoliubov transformation of Eq.~\eref{Bogoliubov_delta_c}, it is possible to find the quasi-particle vacuum-state $\left|\emptyset\right\rangle _{qp}$ in terms of the atomic degrees of freedom
\begin{eqnarray}
  \left|\emptyset\right\rangle _{qp}&=&	\prod_{k}\frac{e^{-\frac{1}{2}e^{i\left(\alpha_{k}+\beta_{k}\right)}\tanh\Lambda_{k}\delta\phi_{k}^{\dagger}\delta\phi_{-k}^{\dagger}}}{\cosh\Lambda_{k}}\left|\emptyset\right\rangle _{a}
\nonumber  \\
&=&	\exp\sum_{k}\left(-\frac{1}{2}e^{i\left(\alpha_{k}+\beta_{k}\right)}\tanh\Lambda_{k}\delta\phi_{k}^{\dagger}\delta\phi_{-k}^{\dagger}-\ln\cosh\Lambda_{k}\right)\left|\emptyset\right\rangle _{a}\, , \label{exponential_operator_vacuum}
\end{eqnarray}
where $\left|\emptyset\right\rangle _{a}$ should be interpreted as the vacuum of excited atoms.

From Eq.~\eref{exponential_operator_vacuum} it is clear that, in the basis of atom occupation number, 
the quasi-particle vacuum is a complicated superposition of states with different number of atoms in the 
condensed $1$-particle state (and a corresponding number of coupled excited atoms, in pairs of opposite momenta). 
Every correlation function is therefore dependent on the entanglement of this many-body atomic state. 

This feature is enhanced by the dynamics,
as can be observed from the scattering operator in Eq.~\eref{Explicit scattering operator} relating early and late times.
The scattering operator acts moving couples of atoms and
the creation of quasi-particles affects the approximated vacuum states differently depending on their features,
in particular depending on the number of atoms occupying the condensed 1-particle state.
The creation of more pairs modifies further the superposition of the entangled atomic states depending on the total number
of atoms and the initial number of excited atoms. 

We can observe this from the action of the condensed state ladder operator,
which does not commute with the the creation of coupled quasi-particles $c_{k}^{\dagger}c_{-k}^{\dagger}$,
which is described by the combination of the operators $\delta\phi_{k}^{\dagger}\delta\phi_{k}$,
$\delta\phi_{k}^{\dagger}\delta\phi_{-k}^{\dagger}$ and $\delta\phi_{k}\delta\phi_{-k}$.
The ladder operator $a_{0}^{\dagger}$ commutes with the first, but not with the others:
\begin{eqnarray}
\left(\delta\phi_{k}^{\dagger}\delta\phi_{k}\right)a_{0}^{\dagger}&=&	a_{0}^{\dagger}\left(\delta\phi_{k}^{\dagger}\delta\phi_{k}\right)\, ,\\
\left(\delta\phi_{k}^{\dagger}\delta\phi_{-k}^{\dagger}\right)^{n}a_{0}^{\dagger}&=&	a_{0}^{\dagger}\left(\delta\phi_{k}^{\dagger}\delta\phi_{-k}^{\dagger}\right)^{n}\left(\frac{N_{0}+1}{N_{0}+1-2n}\right)^{1/2}\, ,\\
\left(\delta\phi_{k}\delta\phi_{-k}\right)^{n}a_{0}^{\dagger}&=&	a_{0}^{\dagger}\left(\delta\phi_{k}\delta\phi_{-k}\right)^{n}\left(\frac{N_{0}+1}{N_{0}+1+2n}\right)^{1/2}\, .
\end{eqnarray}

The operators $a_{0}$ and $a_{0}^{\dagger}$ do not commute with the number-conserving atomic 
ladder operators, and therefore the creation of couples and the correlation functions, up to any order,
will present corrections of order $1/N$ to the values that could be expected in the usual Bogoliubov 
description. Such corrections appear in correlation functions between quasi-particle operators 
and for correlations between quasi-particles and condensate atoms. This is equivalent to saying 
that a condensed state, which is generally not coherent, will present deviations from the expected 
correlation functions predicted by the Bogoliubov theory due to both the interaction and the features of
the initial state itself (through contributions coming from connected diagrams).

\subsection{Entanglement structure in number-conserving formalism}
Within the Bogoliubov description discussed in Section~\ref{sec:orbitas}, the 
mean-field approximation for the condensate is most adequate for states close to coherence, thus 
allowing a separate analysis for the mean-field. The field operator is split in the mean-field function 
$\left\langle \phi\right\rangle$ and the fluctuation operator $\delta\phi$ which is assumed not to 
affect the mean-field through backreaction. Therefore the states in this picture can be written as
\begin{eqnarray}
\left|\left\langle \phi\right\rangle \right\rangle _{mf}\otimes\left|\delta\phi,\delta\phi^{\dagger}\right\rangle _{a\,Bog}\, ,
\end{eqnarray}
meaning that the state belongs to the product of two Hilbert spaces: the mean-field defined on one and 
the fluctuations on the other, with $\delta\phi$ and $\delta\phi^{\dagger}$ ladder operators acting 
only on the second. The Bogoliubov transformation from atom operators to quasi-particles allows to rewrite the state
as shown in Eq.~\eref{exponential_operator_vacuum}. The transformation only affects its second part
\begin{eqnarray}
\left|\left\langle N\right\rangle \right\rangle _{mf}\otimes\left|\emptyset\right\rangle _{qp\,Bog}	=\left|\left\langle N\right\rangle \right\rangle _{mf}\otimes\sum_{lr}a_{lr}\left|l,r\right\rangle _{a\,Bog} \, .\label{Bogol_superposition}
\end{eqnarray}
With such transformation the condensed part of the state is kept separate from the superposition of 
coupled atoms (which here are denoted $l$ and $r$ for brevity) forming the excited part, a separation 
which is maintained during the evolution in the Bogoliubov description. Also the Bogoliubov 
transformation from early-times quasi-particles to late-times quasi-particles affects only the second part
\begin{eqnarray}
\left|\left\langle N\right\rangle \right\rangle _{mf}\otimes\sum_{lr}a_{lr}\left|l,r\right\rangle _{a\,Bog}\Rightarrow\left|\left\langle N\right\rangle \right\rangle _{mf}\otimes\sum_{lr}a_{lr}^{\prime}\left|l,r\right\rangle _{a\,Bog} \, .	 \label{Bogol_evolution}
\end{eqnarray}
In the number-conserving framework there is not such a splitting of the Fock space, and 
there is no separation between the two parts of the 
state. In this case the best approximation for the quasi–particle vacuum is given by a superposition 
of coupled excitations of the atom operators, but the total number of atoms cannot be factored out:
\begin{eqnarray}
\left|N;\emptyset\right\rangle _{qp}	\approx\sum_{lr}a_{lr}\left|N-l-r,l,r\right\rangle _{a}\, . \label{atomic_qp_vacuum}
\end{eqnarray}
The term in the RHS is a superposition of states with $N$ total atoms, of which $N-l-r$ are in the 
condensed 1-particle state and the others occupy excited atomic states and are coupled with each other 
analogously to the previous Eq.~\eref{Bogol_superposition} (the difference being the truncation of the sum, required for a 
sufficiently large number of excited atoms, implying a different normalization). 

The evolution does not split the Hilbert space, and the final state will be a different superposition of atomic states
\begin{eqnarray}
  \sum_{lr}a_{lr}\left|N,l,r\right\rangle _{a}\Rightarrow\sum_{lr}a_{lr}^{\prime}\left(1+O\left(N^{-1}\right)\right)
  \left|N-l-r,l,r\right\rangle _{a}\, .
\end{eqnarray}
We remark that in the RHS the final state must include corrections of order $1/N$ with respect to the 
Bogoliubov prediction, due to the fully quantum behaviour of the condensate ladder operators. These 
are small corrections, but we expect that the difference from the Bogoliubov prediction will be relevant when considering 
many-point correlation functions.

Moreover, these corrections remark the fact that states with different number of atoms in the condensate are transformed 
differently. If we consider a superposition of states of the type in Eq.~\eref{atomic_qp_vacuum} with different total atom
numbers so to reproduce the state in Eq.~\eref{Bogol_superposition}, therefore replicating the splitting of the state, 
we would find that the evolution produces a final state with a different structure, because every state in the 
superposition evolves differently. Therefore, also assuming that the initial state could be written as
\begin{eqnarray}
\sum_{N}\frac{e^{-N/2}}{\sqrt{N!}}\left|N;\emptyset\right\rangle _{qp}\approx
\left|\left\langle N\right\rangle \right\rangle _{mf}\otimes\left|\emptyset\right\rangle _{qp\,Boq}\, ,
\end{eqnarray}
anyway the final state would unavoidably have different features
\begin{eqnarray}
\fl
\sum_{N}\frac{e^{-N/2}}{\sqrt{N!}}\sum_{lr}a_{lr}^{\prime}\left(1+O\left(N^{-1}\right)\right)\left|N-l-r,l,r\right\rangle _{a}
\neq	\left|\left\langle N\right\rangle \right\rangle _{mf}\otimes\sum_{lr}a_{lr}^{\prime}\left|l,r\right\rangle _{a\,Bog} \, .\label{fact}
\end{eqnarray}

We remark that our point is qualitative. Indeed it
is true that also in the weakly
interacting limit the contribution coming from the interaction of Bogoliubov
quasi-particles may be quantitatively larger than the $O\left(N^{-1}\right)$
term in Eq.~\eref{fact}. However, even
if one treats the operator $a_0$ as a number disregarding its quantum
nature, then one cannot have the above discussed entanglement. In that case,
the Hilbert space does not have a sector associated to the
condensed part and no correlation between the condensate and the
quasi-particles is present. To have them one has to
keep the quantum nature of $a_0$, and its contribution to the Hilbert space.

Alternatively, let us suppose to have an interacting theory of bosons for which
no interactions between quasi-particles are present
(one could devise and engineer
similar models). Even in that case one would have a
qualitative difference (and the absence or presence 
of the entanglement structure here discussed) if one retains
or not the quantum nature of $a_0$ and its contribution to the Hilbert space.
Of course one could always argue that in principle
the coupling between the quantum gravity and the matter degrees of freedom
may be such to preserve the factorisation of an initial state.
This is certainly possible in principle,
but it would require a surprisingly high degree of fine tuning at the level of the fundamental theory.

In conclusion, in the Bogoliubov description the state is split in two sectors, and
the total density matrix is therefore a product of two contributions, of which the one relative to the mean-field
can be traced away without affecting the other. The number-conserving picture shows instead that unavoidably
the excited part of the system cannot be manipulated without affecting the condensate.
Tracing away the quantum degrees of freedom of the condensate would imply a loss of information
even without tracing away part of the couples created by analogous curved spacetime dynamics. In other words, 
when one considers the full Hilbert space and the full dynamics, the final
state $\rho_{fin}$ is obtained by an unitary evolution. But now, unlike the
usual Bogoliubov treatment, one can trace
out in $\rho_{fin}$ the condensate degrees of freedom of the Hilbert space,
an operation that we may denote by ``$\Tr_{0}[\cdots]$". So
\begin{equation}
\rho_{fin}^{reduced}=\Tr_{0} [\rho_{fin}]
\end{equation}
is not pure, as a consequence of the presence of the correlations. So one
has $\Tr[\rho_{fin}^2]=1$, at variance with $\Tr[(\rho_{fin}^{reduced})^2] \neq 1$. 
The entanglement between condensate and excited part is an unavoidable feature of the evolution of these states.

\section{Discussion and conclusions}\label{sec:Conc}
The general aim of analogue gravity is to reproduce the phenomenology of quantum field theory 
on curved spacetime with laboratory-viable systems.
In this framework, the geometry is given by a metric tensor assumed 
to be a classical tensor field without quantum degrees of freedom, implying that geometry and 
matter --- the two elements of the system --- are decoupled, \emph{i.e.} the fields belong to distinct Hilbert spaces. 

The usual formulation of analogue gravity in Bose-Einstein condensates reproduces this feature.
In the analogy between the quasi-particle excitations on the condensate and those of scalar quantum 
fields in curved spacetime, the curvature is simulated by the effective acoustic metric derived from the 
classical condensate wavefunction. The condensate wavefunction itself does not belong to the Fock space of 
the excitations, instead it is a distinct classical function.

Moreover, analogue gravity with Bose-Einstein condensates is usually formulated assuming a coherent initial state,
with a formally well defined mean-field function identified with the condensate wavefunction.
The excited part is described by operators obtained translating the atom field by the 
mean-field function and linearising its dynamics; the quasi-particles studied in analogue gravity emerge 
from the resulting Bogoliubov-de Gennes equations.~\footnote{Moreover, 
let us notice that the relation between several quantum gravity scenarios and analogue 
gravity in Bose-Einstein condensates appears to be even stronger than expected,  as in 
many of these models a classical spacetime is recovered by considering an expectation value 
of the geometrical quantum degrees of freedom over a global coherent state the same way 
that the analogue metric is introduced by taking the expectation value of the field on a 
coherent ground state (see e.g.~\cite{Alesci:2019pbs, Gielen:2013naa}).}
The mean-field drives the evolution of the quasi-particles, which have a negligible backreaction on the condensate, and can be 
assumed to evolve independently, in accordance with the Gross-Pitaevskii equation. Neglecting the quantum
nature of the operator creating particles in the condensate one has anyway a unitary dynamics occurring
in the Hilbert space of non-condensed atoms (or, equivalently, of the quasi-particles). 

However, one could still expect an information loss problem to arise whenever 
simulating an analogue black hole system entailing the complete loss of the ingoing Hawking partners, \emph{e.g.} by having a flow with  a region where the hydrodynamical description at the base of the standard analogue gravity formalism fails.
The point is that the Bogoliubov theory is not exact as much as quantum field theory in curved spacetime
is not a full description of quantum gravitational and matter degrees of freedom.

Within the  number-conserving formalism, analogue gravity provides the possibility to develop a more complete description in which one is forced to retain the quantum nature of the 
operator creating particles in the condensate. While this is not \emph{per se} a quantum 
gravity analogue (in the sense that it cannot reproduce the full dynamical equations of the quantum system),
it does provide a proxy for monitoring the possible development 
of entanglement between gravitational and matter degrees of freedom.

It has already been conjectured in quantum gravity that degrees of freedom hidden from the classical
spacetime description, but correlated to matter fields, are necessary to maintain unitarity in the global evolution 
and prevent the information loss~\cite{Perez:2014xca}. 
In order to address the question of whether particle production induces entanglement
between gravitation and matter degrees of freedom,
we have carefully investigated the number-conserving formalism
and studied the simpler process of cosmological particle production in analogue gravity, 
realised by varying the coupling constant from an initial value to a final one. 
We verified that one has a structure of quasi-particles, whose operators
now depend on the operator $a_0$ destroying a particle
in the natural orbital associated to the largest, macroscopic eigenvalue
of the 1-point correlation functions. 

We have shown that also in the number-conserving formalism one can define a
unitary scattering operator, and thus the Bogoliubov transformation from 
early-times to late-times quasi-particles. 
The scattering operator provided in~\eref{Explicit scattering operator}
not only shows the nature of quasi-particle creation as a squeezing process of
the initial quasi-particle vacuum, but also that the evolution process as a whole is unitary precisely
because it entangles the quasi-particles with the condensate atoms constituting the geometry over which the former propagate.

The correlation between the quasi-particles and the condensate atoms
is a general feature, it is not realised just in a regime of high energies --- analogous to the late stages of a black hole
evaporation process or to sudden cosmological expansion --- but it happens during all the evolution~
\footnote{Indeed, the transplanckian problem in black hole radiation may suggest that Hawking quanta might always probe the fundamental degrees of freedom of the underlying the geometry.}, albeit they are suppressed in the number of atoms, $N$, relevant for the system and are hence generally negligible.
When describing the full Fock space there is not unitarity breaking, and the purity of the state is 
preserved: it is not retrieved at late times, nor it is spoiled in the transient of the evolution. 
Nonetheless, such a state after particle production will not factorise into the product of two states --- a condensate (geometrical) and quasi-particle (matter) one --- but, as we have seen, it will be necessarily an entangled state.
This implies, as we have discussed at the end of the previous Section, that an observer unable to access the condensate (geometrical) quantum degrees of freedom would define a reduced density matrix (obtained by tracing over the latter) which would no more be compatible with an unitary evolution.

In practice, in cases such as the cosmological particle creation, where the phenomenon happens on the whole spacetime, $N$ is the (large) number of atoms in the whole condensate and hence the correlations between the substratum and the quasi-particles are negligible. Hence, the number-conserving formalism or the Bogoliubov one in this case may be practically equivalent.  In the black hole case a finite region of spacetime is associated to the particle creation, hence $N$ is not only finite but decreases as a consequence of the evaporation making the correlators between geometry and Hawking quanta more and more non-negligible in the limit in which one simulates a black hole at late stages of its evaporation. This implies that tracing over the quantum geometry degrees of freedom could lead to non-negligible violation of unitary even for regular black hole geometries ({\em i.e.} for geometries without inner singularities, see~{\em e.g.}~\cite{Frolov:1981mz, Hayward:2005gi, Carballo-Rubio:2018pmi}). 

The Bogoliubov limit corresponds to taking the quantum degrees of freedom of the geometry as classical. 
This is not \emph{per se} a unitarity violating operation, as it is equivalent to effectively recover the factorisation of the above mentioned state. Indeed the squeezing operator so recovered (which corresponds to the one describing particle creation on a classical spacetime) is unitarity preserving.  However, the two descriptions are no longer practically equivalent when a region of quantum gravitational evolution is somehow simulated. In this case, having the possibility of tracking the quantum degrees of freedom underlying the background enables to describe the full evolution, while in the analogue of quantum field theory in curved spacetime a trace over the ingoing Hawking quanta is necessary with the usual problematic implications for the preservation of unitary evolution.

In the analogue gravity  picture, the above alternatives would correspond to the fact that the number-conserving evolution can keep track of the establishment of correlations between the atoms and the quasi-particles that cannot be accounted for in the standard Bogoliubov framework. Hence, this analogy naturally leads to the conjecture that a full quantum gravitational description of a black hole creation and evaporation would leave not just a thermal bath on a Minkowski spacetime but rather a highly entangled state between gravitational and matter quantum degrees of freedom corresponding to the same classical geometry.~\footnote{With the possible exception of the enucleation of a disconnected baby universe which would lead to a sort of trivial information loss.}
A very complex state, but nonetheless a state that can be obtained from the initial one (for gravity and matter) via a unitary evolution.

In conclusion, the here presented investigation strongly suggests that
the problems of unitarity breaking and information loss encountered in quantum field theory 
on curved spacetimes can only be addressed in a full quantum gravity description able to keep track of the correlations between quantum matter 
fields and geometrical quantum degrees  of freedom developed via particle creation from the vacuum; 
these degrees of freedom are normally concealed by the assumption of a classical spacetime, but underlay it in any quantum gravity scenario.

\vspace{-10pt}
\section*{Acknowledgments} Discussions with R. Parentani, M. Visser and S. Weinfurtner are gratefully
acknowledged.

\vspace{-10pt}
\section*{References}

\bibliographystyle{hunsrt}
\bibliography{BEC_refs}

\end{document}